\documentclass[aps,prb,floatfix,twocolumn,footinbib]{revtex4-1}

\usepackage{epsfig}
\usepackage{bm}% bold math
\usepackage{amssymb}
\usepackage{graphicx}
\usepackage{epstopdf}
\usepackage{amsmath}
\usepackage{natbib}
\usepackage{hyperref}

\begin{document}

\title{Fano-shaped impurity spectral density, electric-field-induced in-gap state and local magnetic moment  of an adatom on   trilayer graphene}
\author{Zu-Quan Zhang}
\affiliation{School of Physics and Wuhan National High Magnetic field center,
Huazhong University of Science and Technology, Wuhan 430074,  China}
\author{Shuai Li}
\affiliation{School of Physics and Wuhan National High Magnetic field center,
Huazhong University of Science and Technology, Wuhan 430074,  China}
\author{Jing-Tao L\"{u} }
\email{jtlu@hust.edu.cn}
\affiliation{School of Physics and Wuhan National High Magnetic field center,
Huazhong University of Science and Technology, Wuhan 430074,  China}
\author{Jin-Hua Gao}
\email{jinhua@hust.edu.cn}
\affiliation{School of Physics and Wuhan National High Magnetic field center,
Huazhong University of Science and Technology, Wuhan 430074,  China}

%\author{}
%\affiliation{}

\begin{abstract}
Recently, the existence of local magnetic moment in a hydrogen adatom on graphene has been confirmed experimentally [Gonz\'{a}lez-Herrero \emph{et al.}, \emph{Science}, 2016, \textbf{352}, 437]. Inspired by this breakthrough, we theoretically investigate the top-site adatom on trilayer graphene (TLG) by solving the Anderson impurity model via self-consistent mean field method. The influence of the stacking order, the adsorption site and external electric field  are carefully considered. We find that, due to its unique electronic structure, the situation of the TLG is drastically different from that of the monolayer graphene. Firstly,  the adatom on rhombohedral stacked TLG (r-TLG) can have a Fano-shaped impurity spectral density, instead of the normal Lorentzian-like one, when the impurity level is around the Fermi level.  Secondly, the impurity level of the adatom on r-TLG can be tuned into an in-gap state  by an external electric field, which strongly depends on the direction of the applied electric field and can significantly affect the local magnetic moment formation. Finally, we systematically calculate the impurity magnetic phase diagrams, considering various stacking orders, adsorption sites, doping and electric field. We show that, because of the  in-gap state, the impurity magnetic phase of r-TLG will obviously depend on the direction of the applied electric field as well. All our theoretical results can be readily tested in experiment, and may give a comprehensive understanding about the local magnetic moment of adatom  on TLG.
\end{abstract}
\maketitle
\section{introduction}
The adatoms on graphene is an intriguing issue, which  has drawn lots of research interest in the last decade\cite{chemreview2009,adatomreview,chenjh2008,solid2012, prl2011, katsnelson2010,wu2011,hujun2012,jianghua2012,hydrogenSO, zheng2015,wuxiaosong2015,indium2015, prl2008, uchoanjp, hydrogentransport,jafarijpcm, science2016,kondoreview,bulla2010,lilinnjp,zhuanghb2009,uchoakondo2011}. The adatom introduces many important physics to the graphene system, e.g. the modification of transport property\cite{chenjh2008,katsnelson2010,prl2011,solid2012}, the enhancement of spin-orbit interaction\cite{wu2011,hujun2012,jianghua2012,hydrogenSO,zheng2015,wuxiaosong2015,indium2015},  the formation of local magnetic moment\cite{prl2008,uchoanjp, hydrogentransport, jafarijpcm, science2016}, and the Kondo physics\cite{kondoreview,bulla2010,lilinnjp,zhuanghb2009,uchoakondo2011}. One most recent breakthrough is about the hydrogen atom absorbed on graphene, where the local magnetic moment of the hydrogen adatom is confirmed and manipulated in a scanning tunnelling microscopy (STM) experiment\cite{science2016}, nearly ten years after its theoretical proposal. The importance is that a non-magnetic adatom can induce a local magnetic moment on graphene in a controllable way, which can integrate the ferromagnetism into the graphene physics and has great potential in future device applications.

Theoretically, the adatom on graphene can be well described by the Anderson impurity model, where the adatom is viewed as impurity levels coupled to the conducting electrons in the metal (i.e. graphene here)\cite{anderson61}. Due to this hybridization, the discrete impurity spectral is broadened to a Lorentzian-like shape.  And, considering the on-site Coulomb interaction of the impurity orbital,  the energy levels of the spin up and down are separated , which can give rise to the formation of  local magnetic moment. Because of  the novel electronic structure of graphene, the Anderson impurity on graphene has some unique features distinct from the normal metal.  For example, since the density of states (DOS) near Fermi level goes to zero, the effective hybridization between the impurity level and the graphene is very small, which prefers to form a local magnetic moment.

%Impurity atoms in host metal is a fundamental problem in condensed matter physics, which relates to many important phenomena, such as the transport property of metal, the formation of local magnetic moment and the Kondo effect.  %Considering a single impurity, it can be well described by the Anderson model\cite{anderson61}, where the localized impurity orbital is coupled to the conducting electrons of the metal. Due to this hybridization, the impurity %spectral function is broadened from a Delta function to a Lorentzian function. Given the on-site Coulomb interaction of the impurity orbital, the energy levels of the spin up and down are separated , which can give rise to the %formation of the local magnetic moment.

 %The above pictures are also suitable to the adatoms on metal surface, where the line shape of the impurity density of states (DOS) can be precisely measured by the scanning tunnelling microscopy (STM). An intriguing example is %the hydrogen adatom absorbed on graphene, where the spin splitting and local magnetic moment of the hydrogen atom is confirmed by the most recent STM experiment\cite{Herrero437}. On the graphene monolayer, a nonmagnetic adatom %can induce local magnetic moment, which could be manipulated by an external electric field. This novel property has drawn lots of theoretical\cite{prl2008} and experimental interest in the last few years.

 In comparison with the monolayer graphene, trilayer graphene (TLG) has a drastically different electronic structure\cite{rmpgraphene,guinea2006,aba2006,solidtrilayer}. Importantly, with different stacking orders, the corresponding energy dispersion of TLGs are  different.
For example, the rhombohedral (ABC) stacked TLG (r-TLG) has a $\mathbf{k}^3$ dispersion near the Fermi level, and thus it has a  divergent DOS at the Fermi level. Meanwhile, near the Fermi level, the bands of the Bernal (ABA) stacked TLG (b-TLG) are linear or quadratic, and the DOS is finite. Their responses to the electric field are also different. A band gap will be opened by an external electric field in r-TLG, while there is an electric field induced band overlap in the b-TLG\cite{abakoshino,wutrilayer,zhangfantrilayer,koshino2010,lujingtrilayer,heinz2011,stmtrilayergap,abctransport,helin2015}.

Stimulated by the recent experimental progress and considering the novel electronic properties of TLG, in this work, we systematically examine the behaviors of the adatoms on TLG systems  by studying the Anderson impurity model via self-consistent mean field method. We carefully investigate the influences of stacking order, adsorption site, and external electric field, to which little attention has  been paid in the literatures  so far\cite{ding2009}. We find that the adatom on TLG has  some unusual  characteristics, which is distinct from the monolayer graphene as well as the normal metal.

 Firstly,  we find that the adatom on r-TLG can have a Fano-shaped impurity spectral density, in contrast to the common Lorentzian-like line shape.  It actually  results from the interference between the divergent TLG DOS peak at Fermi level and the broadened impurity level. According to our analysis, we argue that it may be general  for any impurity level near the metal DOS singularity.

  Secondly, we demonstrate that a perpendicular electric field can make the impurity level in r-TLG into an in-gap state, which strongly depends on the direction of the applied electric field.
The key issue is that, in addition to opening an energy gap,  the electric field can also shift the energy level of the adatom if it is only coupled to the top layer.
   Namely, with an electric field in one direction, an energy gap can be opened in the r-TLG and the impurity level is shifted into the gap to form an in-gap state. But, when reversing the direction of the electric field,
the impurity level is shifted away and the in-gap state can not be achieved, though the gap can still be opened. This phenomenon appears as long as the adatom is only coupled to the top layer. The in-gap state favors the formation of local magnetic moment, and greatly modifies the impurity magnetic phase diagram of the r-TLG.

Thirdly, we numerically calculate the impurity magnetic phase diagrams for TLG, where various stacking orders, adsorption sites, doping and electric field are considered. We focus on the mechanisms for tuning the adatom magnetic moment via an electric field.
 The key quantity here is the hybridization between the impurity level and the conducting electrons,  which is controllable by an applied electric field.
 There are several typical cases.  For the undoped b-TLG, the electric field induced band overlap  will enhance the hybridization, so that the magnetic phase is suppressed by the electric field.  Meanwhile, for the undoped r-TLG,  the  DOS singularity at the Fermi level results in a giant hybridization, and thus the adatom prefers to be non-magnetic. But, if an electric field is applied to open an energy gap, the hybridization is reduced and thus the adatom on the r-TLG can be tuned to be magnetic.  Importantly, due to the appearance of the in-gap state, the effect of the electric field on the impurity magnetic moment in r-TLG strongly depends on the direction of electric field (i.e. the polarity of the applied bias).
 We also find that the impurity magnetic moment is  very sensitive to the adsorption site. The influence of doping on the impurity magnetic phase diagram on TLG has been discussed as well.

 Considering the rapid experimental progress, we would like to point out that our results here can be readily tested in the STM experiment, and may have potential applications in future magnetic device on graphene systems. It should also be noted that, here we just focus on the top-site adatom, e.g. the case of hydrogen atom, and the cases for hollow and bridge sites are left to a separate work in future.

The rest of the paper is organized as follows: In Sec. \ref{model}, we give our theoretical model and present the formula in the calculation.   In Sec. \ref{discussion}, we first discuss the Fano-shaped impurity spectral function on r-TLG. Then,we talk about the electric field induced in-gap state. Finally,  we show the results about the impurity magnetic phase diagrams of the TLG, and related discussions are also given. A short summary is given in Sec. \ref{summary}.

\section{theoretical model and method}\label{model}
Let us first introduce the theoretical model used in this work.
We use the Anderson impurity model to describe the adatom on the TLG. The complete Hamiltonian is written as
\begin{equation}
H=H_{\textrm{TLG}} + H_{\textrm{imp}} + H_{\textrm{hyb}}.
\end{equation}
Here, $H_{\textrm{TLG}}$ is tight binding Hamiltonian of the TLG, $H_{\textrm{imp}}$ is the Hamiltonian of adatom and $H_{\textrm{hyb}}$ describes the hybridization between the impurity levels and the conducting electrons in TLG.

\begin{figure}[h]
\centering
\includegraphics[width=8cm]{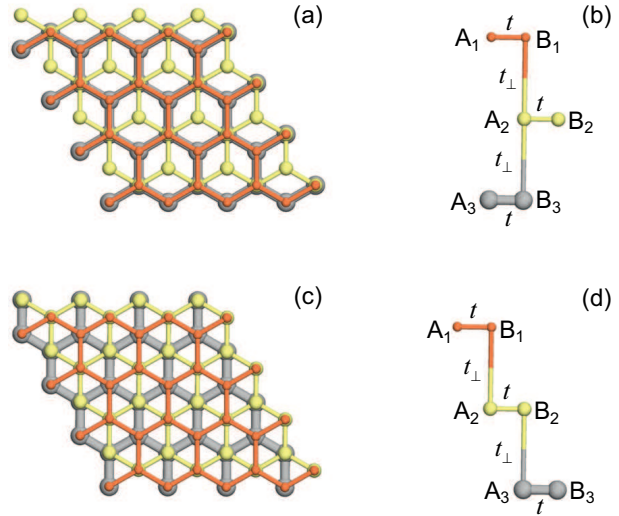}
\caption{(Color online) Schematic diagrams of TLG. (a) and (b): top and side views of b-TLG;  (c) and (d): top and side views of r-TLG.
}
\label{tlg}
\end{figure}
\begin{figure}[h]
\centering
\includegraphics[width=8.5cm]{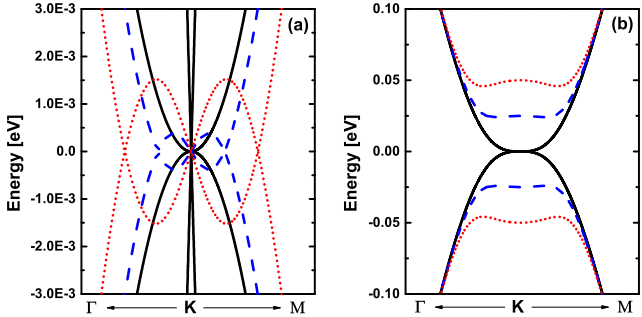}
\caption{(Color online) Energy bands for (a) b-TLG and (b) r-TLG with different bias voltage $V_g$. Solid black lines, $V_{g} = 0 \; \textrm{eV} $; dashed blue lines, $V_{g} = 0.05 \; \textrm{eV} $; dotted red lines, $V_{g} = 0.1 \; \textrm{eV} $.
}
\label{bands}
\end{figure}

For simplicity, only the nearest neighbourhood (NN) hoppings are considered, and the structures of r-TLG and b-TLG are shown in Fig. \ref{tlg}. Generally,
\begin{equation}\label{eqoftlg}
H_{\textrm{TLG}} = H_{\textrm{intra}} + H_{\textrm{inter}} + H_{\textrm{bias}}.
\end{equation}
$H_{\textrm{intra}}$ describes the intralayer hopping, and is equivalent to the Hamiltonian of monolayer graphene
\begin{equation}
H_{\textrm{intra}}=- t \sum_{l\bm{k}\sigma} [ a_{l\bm{k}  \sigma}^{\dag} \phi (\bm{k}) b_{l\bm{k}  \sigma} + h.c.].
\end{equation}
Here, $a_{l\bm{k}\sigma}$ ($b_{l\bm{k}\sigma}$) is the  annihilation operator of the electron on A (B) sublattice, $l$, $\bm{k}$ and $\sigma$ are the layer, momentum, and spin index, respectively, $t$ is the intralayer NN hopping,  $ \phi (\bm{k}) = \sum_{j=1}^{3} e^{i \bm{k} \cdot \bm{\delta}_{j} } $, and $\bm{\delta}_j$ are the three  vectors pointing from one carbon atom on A sublattice  to the three adjacent atoms on  B sublattice in the same layer. As shown in Fig. \ref{tlg}, the interlayer hoppings for the r-TLG and b-TLG are:
\begin{equation} \label{KineInterRTLG}
H_{\textrm{inter}}^{\textrm{r}} = t_{\perp} \sum_{\bm{k}  \sigma}( b_{1\bm{k}  \sigma}^{\dag}  a_{2\bm{k}  \sigma} + b_{2\bm{k}  \sigma}^{\dag}  a_{3\bm{k}  \sigma} + h.c.),
\end{equation}
\begin{equation} \label{KineInterBTLG}
H_{\textrm{inter}}^{b} = t_{\perp} \sum_{\bm{k}  \sigma}( b_{1\bm{k}  \sigma}^{\dag}  a_{2 \bm{k}  \sigma} + a_{2 \bm{k}  \sigma}^{\dag}  b_{3 \bm{k}  \sigma} + h.c.).
\end{equation}
$t_\perp$ is the interlayer hopping parameter.
When a perpendicular  electric field   is applied, it gives
\begin{equation}
H_{\textrm{bias}}=\sum_{l \bm{k} \sigma}\frac{1}{2} ( l - 2 ) V_{g} ( a_{l \bm{k} \sigma}^{\dag} a_{l \bm{k} \sigma}+ b_{l \bm{k} \sigma}^{\dag} b_{l \bm{k} \sigma}).
\end{equation}
$V_g$ is the applied bias voltage, i.e. the potential difference between the top and bottom layers. Diagonalizing the Hamiltonian of the TLG in Eq. (\ref{eqoftlg}) , we can  get the corresponding energy bands
\begin{equation}
H_{\textrm{TLG}}=\sum_{n\bm{k}\sigma} (\varepsilon_{n\bm{k}\sigma}-\mu)c^{\dag}_{n\bm{k}\sigma}c_{n\bm{k}\sigma}
\end{equation}
where $n$ is the band index and $c^{\dag}_{n\bm{k}\sigma}$ is the electron creation operator of the eigenstate. The chemical potential $\mu$ is also included in order to consider the charge doping.
In Fig. \ref{bands}, we show the energy bands for both r-TLG and b-TLG with different bias voltages  as an example, which is helpful to understand the following numerical results. The electric field induced band overlap of the b-TLG, as well as  the opened gap of r-TLG,  are shown clearly in Fig. \ref{bands}. In this paper, we set $t=3.16$ eV, and $t_\perp = 0.39$ eV.  Note that, in experiment, one  way to produce a perpendicular electric field is to build a dual-gated device\cite{zhangyuanbo2009,lauaba,lau2014}, where the potential difference $V_g$ and the charge density (or chemical potential $\mu$) can be independently tuned. Another possible way is  to use molecular doping\cite{dopinggap,duan2011,wuxiaosong}, which not only can induce a potential difference between layers, but also change the charge density. Here, we assume that the electric field is produced by the dual-gated device, where the impurity level is shifted in the presence of the electric field.

The adatom is described as a single level impurity with the on-site Coulomb interaction. Note that in this work, we assume that the adatom is on the top layer ($l=1$). Thus, in reality, the energy level of the impurity $\varepsilon_0$ can also be shifted by the external electric field\cite{levelshift}. The impurity Hamiltonian is
\begin{equation}
H_{\textrm{imp}} =  \sum_{\sigma} \varepsilon_d  d_{\sigma}^{\dag} d_{\sigma} + U  d_{\uparrow}^{\dag} d_{\uparrow}  d_{\downarrow}^{\dag} d_{\downarrow}.
\end{equation}
where $\varepsilon_d =  (\varepsilon_{0}- \frac{V_g}{2})$ is the shifted energy of the impurity level by the electric field.  $U$ is the on-site Coulomb interaction.

%Here, we consider three different absorbing sites. Two of them are top sites, i.e. the adatom is on the top of the $\textrm{A}_1$ or $\textrm{B}_1$ carbon atom of the top layer, as shown in Fig. \ref{tlg}. Note that in TLG, the A and B sublattice in one monolayer are no longer equivalent. The hybridization of the top site impurity are

As mentioned above, we only consider the top-site adatom, i.e. the case of hydrogen atom.
So, as shown in Fig. \ref{tlg}, the adatom is on the top of the $\textrm{A}_1$ or $\textrm{B}_1$ carbon atom of the top layer. Note that in TLG, the A and B sublattice in one monolayer are no longer equivalent. The hybridization of the top site impurity are

\begin{equation}
H_{\textrm{hyb}}^{\textrm{A}_1} = \frac{V}{\sqrt{N}} \sum_{\bm{k}  \sigma} ( d_{\sigma}^{\dag} a_{1 \bm{k} \sigma} + h.c. ),
\end{equation}
\begin{equation}
H_{\textrm{hyb}}^{\textrm{B}_1} = \frac{V}{\sqrt{N}} \sum_{\bm{k}  \sigma} ( d_{\sigma}^{\dag} b_{1 \bm{k} \sigma} + h.c. ).
\end{equation}
Here, $N$ is the number of sites on sublattice A or B in one monolayer.

%Here, the impurity atom at the hollow site will hybridize with the six adjacent carbon atoms, and the corresponding hybridization Hamiltonian is
%\begin{equation}
%H_{\textrm{hyb}}^{h} = \frac{V}{\sqrt{N}} \sum_{\bm{k}  \sigma} [ d_{\sigma}^{\dag} \phi (\bm{k}) a_{1 k \sigma} + d_{\sigma}^{\dag} {\phi}^{\ast} (\bm{k}) b_{1 k \sigma} + h.c. ]
%\end{equation}

With the mean field approximation, the occupation of the impurity level is given by
\begin{equation}\label{nd}
\begin{split}
 n_{\sigma} &=   \int_{- \infty}^{\mu} \textrm{d} \omega \rho_{d,\sigma}(\omega)   \\
                          &= - \frac{1}{\pi} \int_{- \infty}^{\mu} \textrm{d} \omega \textrm{Im} [ G_{dd, \sigma}^{r}(\omega) ] ,
\end{split}
\end{equation}
where $\rho_{d,\sigma}(\omega)$ is the impurity spectral density, and $G_{dd, \sigma}^r(\omega)$ is the retarded Green's function of the impurity electron. $G_{dd, \sigma}^r(\omega)$ can be expressed by the impurity self-energy
\begin{equation}\label{greenf}
G_{dd, \sigma}^r(\omega) =\frac{1}{\omega + i \eta - \varepsilon_{d} - U  n_{\bar{\sigma}} - \Sigma_{dd}^{r}(\omega) }.
\end{equation}
When the adatom is on the top site, e.g. on $\textrm{A}_1$ site, the impurity self-energy is
\begin{equation}
\begin{split}\label{selfenergy}
\Sigma_{dd}^r(\omega) &=(V^2/N) \sum_{\bm{k}} g^{r}_{a_1a_1,\sigma}(\bm{k},\omega)  \\
                      &=V^2 g^r_{a_1 a_1, \sigma}(\omega)
\end{split}
\end{equation}
where $g^{r}_{a_1a_1,\sigma}(\bm{k},t)=-i\theta(t) \langle \{a_{1\bm{k}\sigma}(t),a^\dagger_{1\bm{k}\sigma}(0)\} \rangle$ is the noninteracting Green's function of the electrons in TLG, and we define
\begin{equation}
g^r_{a_1a_1,\sigma}(\omega)\equiv (1/N)\sum_{\bm{k}} g^{r}_{a_1a_1,\sigma}(\bm{k},\omega).
\end{equation}
We see that $g^r_{a_1a_1,\sigma}(\omega)$ is actually  the real space noninteracting Green's function of TLG electron on the $\textrm{A}_1$ site.  Thus, the local density of states (LDOS) of pristine TLG on the $\textrm{A}_1$ site is
\begin{equation}\label{ldos}
\rho^0_{a_1}(\omega)=-\frac{1}{\pi} \textrm{Im}[g^r_{a_1 a_1, \sigma} (\omega)].
\end{equation}
Comparing with Eq. (\ref{selfenergy}), we see that  $\rho^0_{a_1}(\omega)$ is equal to the imaginary part of the self-energy $\Sigma^r_{dd}(\omega)$, apart from a constant. Meanwhile,  for the Anderson impurity, the hybridization function
\begin{equation}
\Gamma(\omega)=\pi V^2 \rho^0_{a_1}(\omega)
\end{equation}
is normally used to describe the  the coupling between the impurity level and the conducting electrons in metal (here, we assume the adsorption site is $\textrm{A}_1$ for example). Thus, the LDOS of the metal at the adsorption site actually reflects the hybridization function. We would like to emphasize that, in normal metal, the electron distribution is uniform, so that the LDOS is  uniform as well and is in some sense equivalent to the DOS. But, here, the A and B sublattices are inequivalent. Thus, the LDOS is now site dependent and is different from the DOS of the whole system. It means that only the LDOS of the adsorption site, instead of the DOS of the whole system,  is meaningful for the Anderson impurity.
%In the Anderson impurity model, we can define a hybridization function
%  \begin{equation}
%\Gamma(\omega)=\pi V^2 \rho^0_{a_1}(\omega)
%\end{equation}
%to describe the coupling between the impurity level and the conducting electrons in metal (here, we assume the adsorption site is $\textrm{A}_1$ for example). The LDOS of the metal at the adsorption site actually is equivalent %to the hybridization function.

Numerically, Eqs.(\ref{nd}-\ref{greenf}) can be self-consistently solved, and then we can get impurity magnetic phase diagram, as well as the impurity spectral function, for all the cases.

\section{results and discussions}\label{discussion}
Here, we present our numerical results and related discussions. First, we would like to discuss the interesting impurity spectral density of the r-TLG, which may be of a Fano line shape instead of the normal Lorentzian-like one. Second, we talk about the  electric field induced in-gap state in r-TLG, which can greatly influence the  magnetic moment of adatom.  Finally, we give the impurity magnetic phase diagrams of the TLG, where the effects of the stacking order, adsorption site, doping and electric field are discussed in detail. We interpret the mechanisms for controlling the impurity magnetic moment by electric field.

\subsection{Fano-shaped impurity spectral density in r-TLG}
\begin{figure}[h]
\centering
\includegraphics[width=8.5cm]{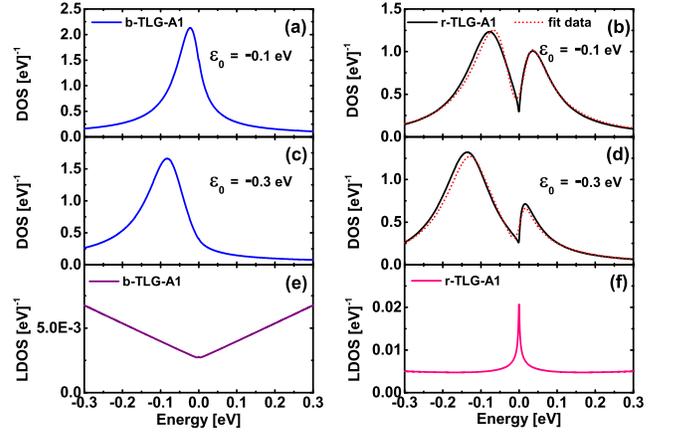}
\caption{(Color online) (a), (b), (c), (d): The impurity spectral density $\rho_d(\omega)$ in the noninteracting case ($U=0$).  For r-TLG [(b) and (d)], the red dotted lines are the fitting curves for the Fano line shape with Eq.(\ref{rhod}). The fitting parameters  $[c_{1}, \varepsilon_{1}, \Gamma_{1}, z_{2}, \varepsilon_{2} ,\Gamma_{2}, q ]$ are [0.51,$-0.032$ $\textrm{eV}$, 0.073 $\textrm{eV}$, 0.39 $\textrm{eV}$, 0.012 $\textrm{eV}$, 0.038 $\textrm{eV}$, 0.35] in (b) and
 [0.49, $-0.10$ $\textrm{eV}$, 0.073 $\textrm{eV}$, 0.75 $\textrm{eV}$, 0.0043 $\textrm{eV}$, 0.020  $\textrm{eV}$, 0.76] in (d).  (e) and (f): The corresponding LDOS of TLG at $A_1$ site.  Other parameters are $V=4$ eV, $V_g=0$ eV, $\mu=0$ eV.
}
\label{fanodos}
\end{figure}

To see the spectral density of the adatom, we first consider the noninteracting case, i.e. $U=0$, where the impurity levels for spin up and down are degenerate. By solving Eq. (\ref{nd}), we plot the impurity spectral $\rho_{d}(\omega)$ in Fig. \ref{fanodos} for the cases that the adatom is on the $\textrm{A}_1$ site.
Obviously, the impurity spectral density of the b-TLG and r-TLG have different behaviors. For the b-TLG, the impurity spectral density coincides with the common understanding. The impurity level is broadened by the metal conducting electrons, which has a Lorentzian-like line shape. The width of the DOS peak depends on the imaginary part of the impurity self-energy, i.e. the LDOS given in Eq. (\ref{ldos}), and the corresponding LDOS is plotted in  Fig. \ref{fanodos} (e). Meanwhile,  the impurity spectral density of b-TLG does not depend much on the position of the impurity level $\varepsilon_0$, as shown in Fig. \ref{fanodos} (a) and (c).

The behaviours of the r-TLG are completely different. In Fig.\ref{fanodos} (b), when the impurity level is around the Fermi level ($\varepsilon_0=-0.1$ eV), the impurity spectral density is not a  peak as normal, but has a dip at the Fermi level.
When $\varepsilon_0$ is shifted away ($\varepsilon_0=-0.3$ eV), it recovers to a Lorentzian-like shape, but has a small peak near the Fermi level, as shown in Fig. \ref{fanodos} (d).

\begin{figure}[h]
\centering
\includegraphics[width=8.5cm]{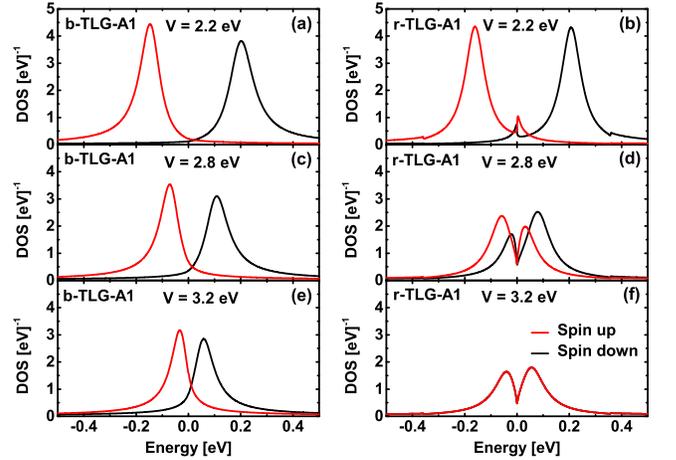}
\caption{(Color online)  The impurity spectral density $\rho_{d,\sigma}(\omega)$ with finite on-site $U$ ($U=0.9$ eV). $\mu=0$ eV, $\varepsilon_{0} = - 0.4 \; \textrm{eV}$.
}
\label{spin}
\end{figure}

We now demonstrate that the unusual impurity spectral density of r-TLG can be interpreted as a Fano resonance, which reflects the interference between the broadened impurity level and the DOS peak of r-TLG at the Fermi level.
As we know, the shape of the  $\rho_{d}(\omega)$ actually depends on the imaginary part of the self-energy, which is nearly a constant in normal metal. But, in r-TLG, this assumption is invalid. We plot the corresponding LDOS, i.e. the imaginary part of the self-energy,  in Fig. \ref{fanodos} (f),  which has a sharp peak at the Fermi level in contrast to the b-TLG case [see Fig. \ref{fanodos} (e)]. This is not surprising because that the DOS of r-TLG is divergent at the Fermi level. To understand the shape of  $\rho_{d}(\omega)$,  intuitively, we can separate $\Sigma^r_{dd}(\omega)$ into two parts by sorting the electron states in the summation in the definition of Eq. (\ref{selfenergy}),
%As mentioned above, the impurity self-energy $\Sigma^r_{dd}(\omega)$ is proportional to $g^r_{a_1a_1,\sigma}(\omega)$, which determines the LDOS on the adsorption site of the pristine TLG. We plot the LDOS $\rho^0_{a_1}(\omega)$ %of the r-TLG in Fig. \ref{fanodos}(e)(f). Intuitively, $\rho^0_{a_1}(\omega)$ consists of two parts: one part is a DOS peak near $E=0$; the other is a constant DOS for all the energy values. Correspondingly, in the definition %Eq. \ref{selfenergy},  we can separate $\Sigma^r_{dd}(\omega)$ into two parts by sorting the electron states in the summation
\begin{equation}
\Sigma^r_{dd}(\omega)=\Sigma^{r(1)}_{dd}(\omega) + \Sigma^{r(2)}_{dd}(\omega).
\end{equation}
$\Sigma^{r(1)}_{dd}(\omega)$ corresponds to the constant background of the LDOS, and $\Sigma^{r(2)}_{dd}(\omega)$ is related to the peak.
Specifically, from the band structure of r-TLG (see in Fig. \ref{bands}), we see that only the states near the $E=0$ contribute to the DOS peak, and thus by summing all these states, we approximately have
\begin{equation}
\Sigma^{r(2)}_{dd}(\omega) \approx V^2 g^r_{\textrm{peak}}(\omega)
\end{equation}
where
\begin{equation}\label{grpeak}
g^r_{\textrm{peak}}(\omega) = \frac{c_{2}}{\omega - \varepsilon_2 + i\Gamma_2}.
\end{equation}
Here, $c_{2}$ is the appropriate strength of the pole, $\varepsilon_2$ is the position of DOS peak ($\varepsilon_2=0$ in this case). We just  use a Lorentzian to model the DOS peak here, and in principle the peak width $\Gamma_2$ should be very narrow (several tens of meV \cite{xiao2015}) in comparison with the  hybridization broadened impurity level (normally several eV). Meanwhile, the sum of all the other states gives $\Sigma^{r(1)}_{dd}(\omega)$. Note that
\begin{equation}\label{gamma1}
\Gamma_1 \equiv -\textrm{Im}[\Sigma^{r(1)}_{dd}(\omega)] \approx {\rm const}.
\end{equation}
which results from the nearly constant LDOS except for the DOS peak, as shown in Fig. \ref{fanodos} (f). Now, the impurity Green's function can be approximately expressed as
\begin{equation}\label{dyson}
G^{r}_{dd} (\omega) \approx G^{0r}_{dd}(\omega) + G^{0r}_{dd}(\omega)V g^r_{\textrm{peak}}(\omega)VG^{0r}_{dd}(\omega),
\end{equation}
where we define
\begin{equation}
G^{0r}_{dd}(\omega) \equiv \frac{c_1}{\omega-\varepsilon_d - \Sigma^{r(1)}_{dd}(\omega) + i\eta}.
\end{equation}
Clearly, the $G^{0r}_{dd}(\omega)$ gives the broadened impurity level,
\begin{equation}
\rho^0_d(\omega)=-\frac{1}{\pi}\textrm{Im}[G^{0r}_{dd}(\omega)],
\end{equation}
which should be a Lorentzian due to the relation (\ref{gamma1}), and the width of the corresponding DOS peak is given by $\Gamma_1$. And the position of the DOS peak here is $\varepsilon_{1} = \varepsilon_{d} + \mathrm{Re}[\Sigma_{dd}^{r(1)}(\omega)]$. Here, $c_1$ is also the  strength of the pole. Finally, by Eq. (\ref{dyson}), we get an approximate expression of the impurity spectral density
\begin{equation}\label{rhod}
\rho_d(\omega) = \rho^{0}_d (\omega) + z_{2} [ \rho^{0}_d(\omega)]^2 \frac{q^2 - 1 + 2 q \varepsilon'}{1+ {\varepsilon'}^2},
\end{equation}
where $z_2=\frac{\pi c_2 V^2}{\Gamma_2}$, $\varepsilon' = \frac{\omega - \varepsilon_2}{\Gamma_2}$ and
\begin{equation}
q=-\frac{\textrm{Re}[G^{0r}_{dd}(\omega) ]}{\textrm{Im}[G^{0r}_{dd}(\omega) ]}
\end{equation}
The factor $(q^2 - 1 + 2 q \varepsilon')/(1+ {\varepsilon'}^2)$ indicates that the impurity spectral density has a Fano profile, which is determined by the parameter q.  We want to point out that the situation here is very similar as the Kondo physics of a magnetic adatom on metal surface\cite{hgluo2004}. The Kondo resonance of a magnetic impurity will give rise to a narrow DOS peak at the Fermi level, which can interference with the broadened impurity level, and thus induce a Fano line shape of the impurity spectral density.  The band structure of r-TLG has a DOS peak at the Fermi level, which actually takes the place of  the Kondo peak in the Fano resonance. The distinction is that the DOS peak here is an intrinsic property of the band structure, so that the Fano-shaped impurity spectral density we predicted here does not need very low temperature, which is required in the Kondo case.

Then, we test the above picture about the Fano resonance by fitting the numerical results of the impurity spectral density of the r-TLG with the formula in Eq. (\ref{rhod}). The results of fitting are given in Fig. \ref{fanodos} (b) and (d) (see the dotted lines). We see that the formula of the Fano resonance can well describe the impurity spectral density. It means that, in Fig. \ref{fanodos} (b), the dip at the Fermi level should come from a Fano resonance between the DOS peak of r-TLG and the broadened impurity level.

When the on-site Coulomb interaction is included, a local magnetic moment can occur with proper values of the hybridization and Coulomb interaction $U$. This is well described by the Anderson impurity model. If the adatom becomes magnetic, the impurity levels for the up spin and down spin are no longer degenerate. In this case,  we have two Fano-shaped impurity spectral density for both spin up and down, as shown in Fig. \ref{spin} (b) and (d).  However, if the impurity is non-magnetic, the spin is degenerate, and only one  impurity spectral density can be observed, see Fig. \ref{spin} (f). The corresponding cases of b-TLG are given in Fig. \ref{spin} (a), (c) and (e) as well.

The results above are all about the adatom on $A_1$ site. Actually, the Fano-shaped impurity spectral density can only be observed in the r-TLG with $A_1$ site adatom. This is because that, when adatom is on $B_1$ site, the corresponding LDOS does not have a peak at the Fermi level, even for the r-TLG.

At last,  we would like to argue that the Fano-shaped impurity spectral density may be a general phenomenon when the impurity level is near a DOS singularity of the metal. Actually, in our analysis above, the only assumption is to use a Lorentzian to model the DOS peak in metal, see in Eq. (\ref{grpeak}). Thus, it is essentially suitable for any DOS singularity in metal. Namely, we predict that, when the metal has a DOS singularity and the impurity level is near the singularity, the impurity DOS should be of a Fano line shape instead of the normal Lorentzian one. We hope this prediction can be tested in future experiment.

\subsection{Electric-field-induced in-gap state  in r-TLG}
The in-gap state has been intensively studied in bilayer graphene\cite{wuxiaosong,ingap2007,vacancy2010,ingap2013,ingap2015,zhujianxin}, which should also be important for the TLG.
Here, we would like to point out that, for the adatom on r-TLG,  a perpendicular electric field has two important effects: shifting the impurity level and opening an energy gap. An intriguing consequence is the electric field induced in-gap state,  which strongly depends on the direction of the electric field, or the polarity of the applied bias.
Assuming that the impurity level is below the Fermi level ($\varepsilon_0- \mu < 0$), a negative bias ($V_g < 0$) will not only open an energy gap but also shift the impurity level towards the gap, since $\varepsilon_d = \varepsilon_0 - V_g /2$. Thus, an in-gap state can be made in this case. However, if the direction of the electric field is reversed ($V_g>0$), a gap can still be opened but the impurity level is now moved away from the gap, so that the in-gap state can never be achieved.

\onecolumngrid

\begin{center}
\begin{figure}[h!]
\includegraphics[width=15cm]{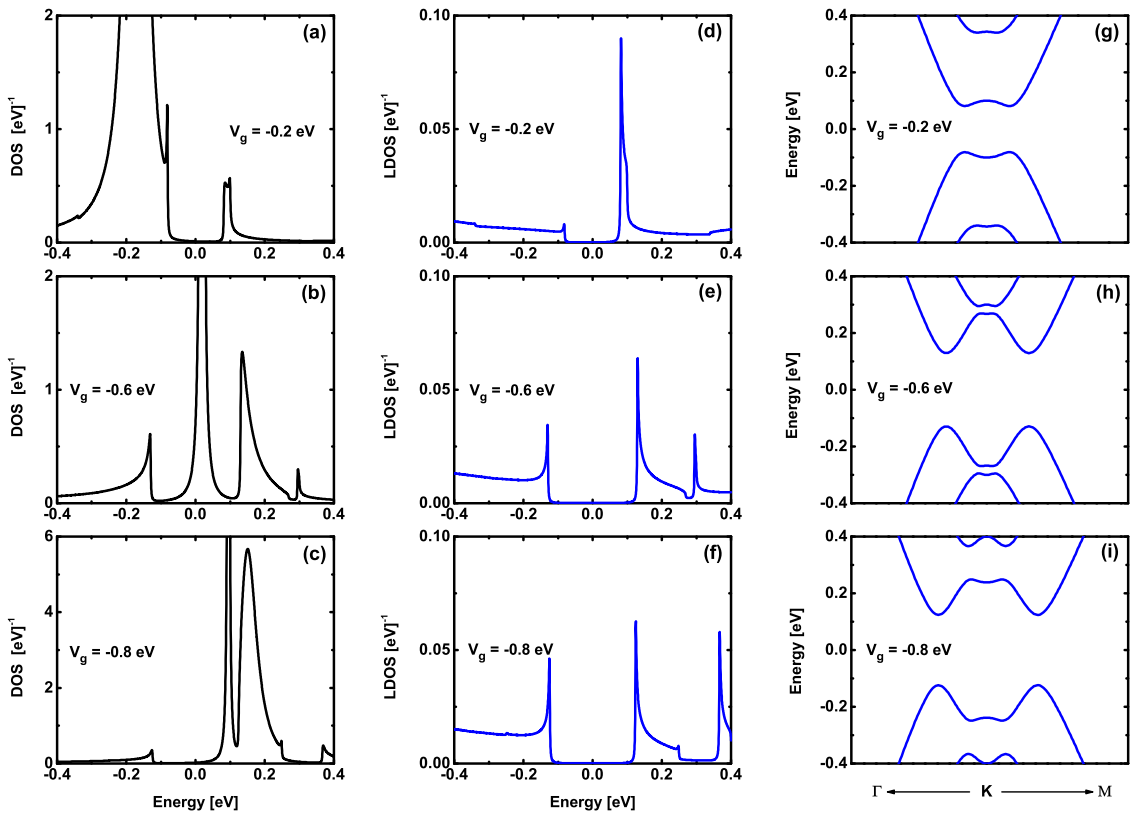}
\caption{(Color online) (a),(b),(c): The impurity spectral density $\rho_d (\omega)$ of the r-TLG-$A_1$ in the noninteracting case with different $V_g$. (d),(e),(f):  The corresponding LDOS of TLG at $A_1$ site.  (g),(h),(i): The corresponding energy bands of the r-TLG. $\varepsilon_{0} = -0.3 \ \textrm{eV}$,  $V = 1 \ \textrm{eV}$.
}
\label{ingapfig}
\end{figure}
\end{center}

\twocolumngrid

The above picture of the electric field induced in-gap state has nothing to do with the on-site $U$ of the adatom.  Thus,  we first give
 a concrete example in the case of $U=0$ in Fig. \ref{ingapfig} (i.e. spin degenerate case). Here, we consider an adatom with $\varepsilon_0 = -0.3$ eV on $A_1$ site, and apply bias voltages $V_g=$ $-0.2$ eV, $-0.6$ eV and $-0.8$ eV, respectively.  When $V_g=-0.2$ eV, an energy gap is opened as shown in Fig. \ref{ingapfig} (g). In Fig. \ref{ingapfig} (d), we plot the corresponding LDOS at $A_1$ site. There is a peak at the edge of the conduction band but no obvious peak for the valence band.
  It is known that, due to the opened gap,  the DOS  has singularities at the edges of both the conduction and valence bands in  r-TLG. But the LDOS at $A_1$ site is asymmetric  because the electrons on $A_1$ site are mainly from the conduction bands.
 In Fig. \ref{ingapfig} (a), we see that the $V_g=-0.2$ eV is not large enough to move the impurity level into the gap, and we observe a broadened impurity level out of the gap. With a larger bias  $V_g=-0.6$ eV, the impurity level can now be shifted into the gap, and an in-gap state appears as shown in Fig. \ref{ingapfig} (b), where the corresponding LDOS and bands are given in Fig. \ref{ingapfig} (e) and (h), respectively.
Increasing $V_g$ further, the impurity level is moved up to the conduction band,  which is the case in Fig. \ref{ingapfig} (c), (f) and (i) with $V_g = -0.8$ eV. It should be noted that, in Fig. \ref{ingapfig} (c), the shifted impurity level is near the edge of conduction band. We actually get a Fano-shaped impurity spectral density, instead of the Lorentzian one. The reason is similar as the case in last section, since LDOS has a sharp DOS peak at the band edge. Reversing the direction of electric field, $V_g$ becomes positive and the impurity level will be pulled down to lower energy. So, we can not get an in-gap state with positive $V_g$.  We indeed do not observe any in-gap state in our numerical results. When the impurity level is above the chemical potential ($\varepsilon_0 - \mu > 0$), the situation  can be treated similarly. And if the adatom is on the $B_1$, an in-gap state can also be achieved by a proper electric field.

\begin{figure}[h!]
\centering
\includegraphics[width=8.5cm]{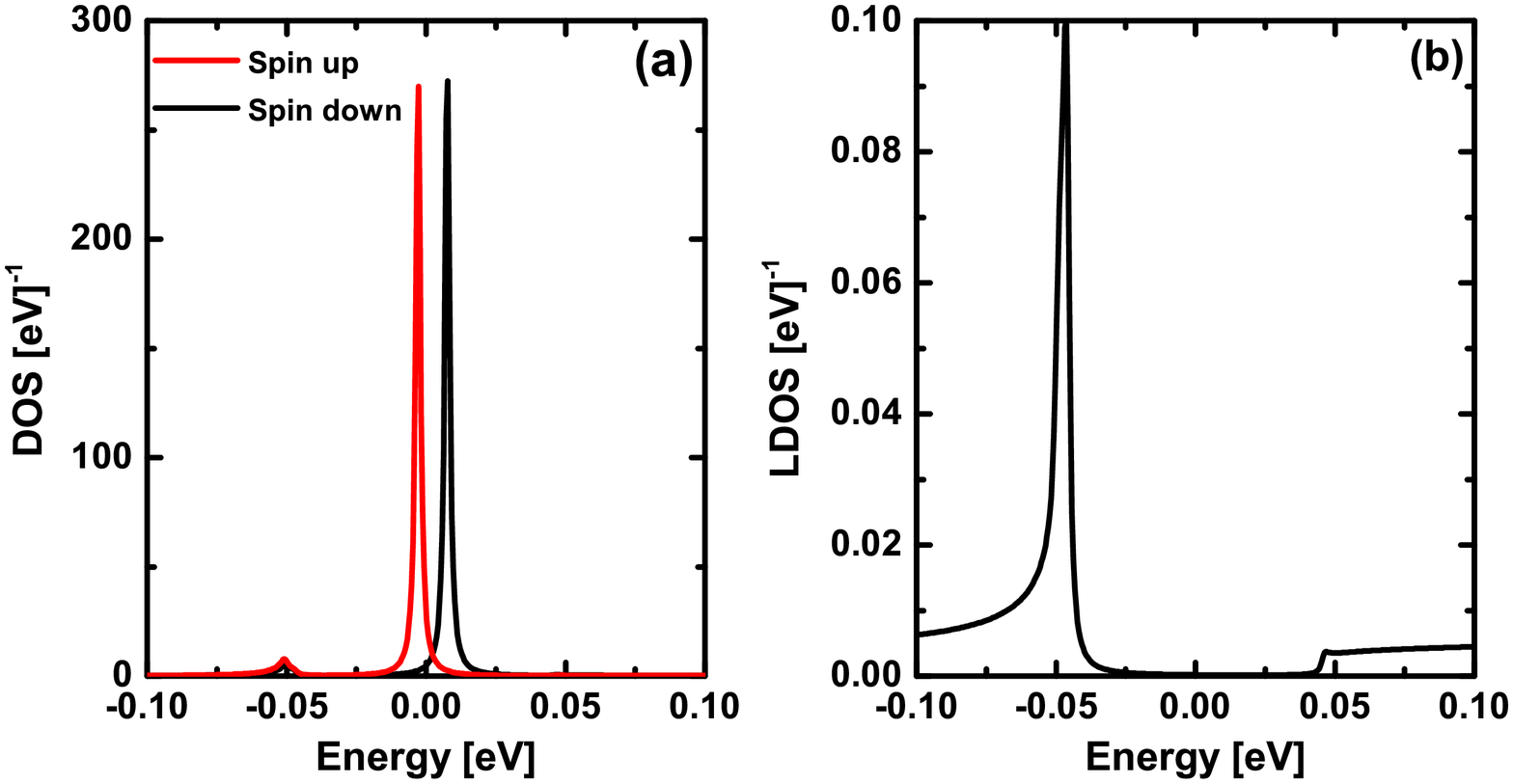}
\caption{(Color online) (a) The impurity spectral density $\rho_{d,\sigma} (\omega)$ of the r-TLG-$A_1$ with  $U = 0.0157 \; \textrm{eV}$. Other parameters are $\mu = 0 \; \textrm{eV}$,  $V = 0.5 \; \textrm{eV}$, $V_{g} = 0.1 \; \textrm{eV}$, $\varepsilon_{0}=-7.85 \times 10^{-3} \; \textrm{eV}$.  The corresponding dimensionless parameters in the magnetic phase diagram  are $x=5$ and $y=0.5$. (b) The corresponding LDOS of TLG at $A_1$ site.
 }
\label{ingapmagnetic}
\end{figure}

 The in-gap state prefers to be magnetic, if the on-site Coulomb interaction is considered.  This is because that , for the in-gap state,  the effective coupling between the impurity level and the conducting electrons is very weak. So, a tiny on-site interaction can break the spin degeneracy, and the adatom becomes magnetic in the undoping case. We give an example in Fig. \ref{ingapmagnetic}, where a small on-site $U$ about $15$  meV can induce a spin splitting for the in-gap state, even if the hybridization constant $V=0.5$ eV  is not small. In the next section, we will show that the electric-field-induced in-gap state will strongly influence the formation and control of the local magnetic moment  of the adatom.

\subsection{Local magnetic moment of the adatom on TLG}
The local magnetic moment formation of an impurity in metal is well described by the Anderson impurity model. The distinct electronic structure of the TLG will make the impurity magnetic phase diagram different from  the case of normal metal. We will show that both the stacking order and the external electric field can obviously modify the local magnetic moment formation on the TLG.

\begin{figure}[h!]
\centering
\includegraphics[width=8.5cm]{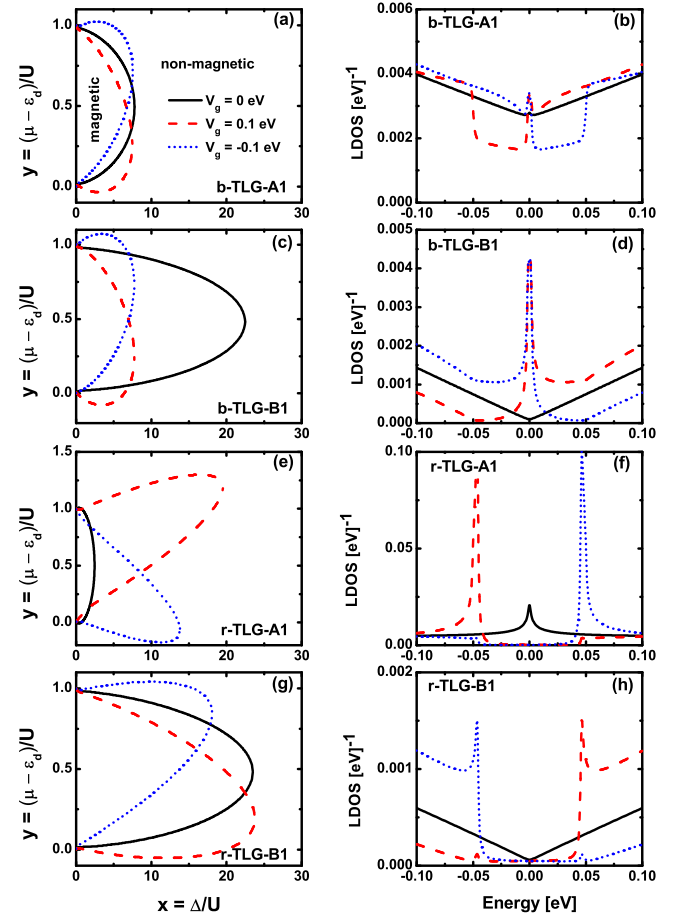}
\caption{(Color online) (a), (c), (e), (g): The magnetic phase boundaries of an impurity on the undoped TLG ($\mu=0$ eV) with the impurity level as a variable. (b),(d),(f),(h): The corresponding LDOS of TLG on the adsorption site.  $V = 0.5 \; \textrm{eV}$, $D=10$ eV.
}
\label{phase1}
\end{figure}

 Let us first introduce some basic physical pictures about the local magnetic moment formation. In the Anderson impurity model, the formation of local magnetic moment depends on the occupation of the impurity level for each spin component $n_\sigma$, which is calculated self-consistently by Eq. (\ref{nd}). The local magnetic moment becomes nonzero when $n_\sigma \neq n_{\bar{\sigma}}$, otherwise the adatom is non-magnetic. Note that here we do not consider the Kondo effect, which is beyond the mean field picture and should occur at very low temperature.  There are several key parameters, which govern the local magnetic moment formation. One is   the on-site Coulomb interaction $U$. The Coulomb interaction favors the formation of local magnetic moment, because it tends to prohibit the double occupation of the impurity level. The  hybridization between the impurity electron and the conducting electrons in metal is also a key quantity, which tends to wash out the local magnetic moment.
 We use the hybridization function $\Gamma(\omega)$, or equivalently the LDOS at the adsorption site of the pristine TLG, to represent the hybridization.
 In normal metal,  the hybridization function can be viewed as a constant $\Delta=\pi V^2 /D$, i.e. the hybridization energy,  because that the DOS in metal normally varies smoothly with the energy and can be approximately viewed as a constant. Here, $D$ is the band width and $1/D$ is a constant DOS. However, due to the unique DOS of TLG, the LDOS should not be considered as a constant any more, and  will play a key role in the formation of the local magnetic moment. Another important quantity is the energy position of the impurity level relative to the Fermi level of the metal. When the impurity level is close to the Fermi level, the system is in the mixed valence region and small hybridization can destroy the local magnetic moment.

In Fig. \ref{phase1}, we show the phase boundaries of the adatom between the magnetic and non-magnetic phases for both r-TLG and b-TLG in the undoped case.
 As is known, the $\textrm{A}_1$ and $\textrm{B}_1$ sites of TLG are no longer equivalent, so that  two adsorption sites are considered here. Specifically, there are four concrete situations in Fig. \ref{phase1}: b-TLG with $\textrm{A}_1$ site adsorption (b-TLG-$\textrm{A}_1$), b-TLG with $\textrm{B}_1$ site adsorption (b-TLG-$\textrm{B}_1$), r-TLG with $\textrm{A}_1$ site adsorption (r-TLG-$\textrm{A}_1$), r-TLG with $\textrm{B}_1$ site adsorption (r-TLG-$\textrm{B}_1$). And in order to illustrate the influence of the external electric field, we  also plot the results with different bias voltage $V_g$.
   In the phase diagram, $x=\Delta/U$  and $y=(\mu-\varepsilon_d)/U$ are two dimensionless parameters, which represent the hybridization and the energy position of the impurity level relative to the chemical potential $\mu$, respectively. The two parameters $x$ and $y$ determine the magnetic phase of the adatom.
Here, we set $\mu=0$ (charge neutrality) and let   $\varepsilon_d$  be variable. It corresponds to the case of the undoped TLG with different impurity energy levels $\varepsilon_0$, since $V_g$  is given.
It should be noted that, to fix  $\mu$ or $\varepsilon_d$  gives different information in the case of TLG, though $y=(\mu-\varepsilon_d)/U$ may be the same. We discuss the case with fixed impurity energy level later.
In order to interpret the numerical results, we  plot the corresponding LDOS in Fig. \ref{phase1} (b), (d), (f) and (h) as well,  which can give the effective hybridization together with the constant hybridization energy $\Delta$.

We now discuss the results without the external electric field ($V_g=0$, see the black solid lines in Fig. \ref{phase1}). First, we see that the magnetic regions in the phase diagrams are different. The magnetic phase of r-TLG-$\textrm{A}_1$ is much smaller than other cases [see Fig. \ref{phase1}(e)], while that of the b-TLG-$B_1$ [see Fig. \ref{phase1} (c)] and r-TLG-$B_1$ [see Fig. \ref{phase1} (g)] are the largest. This is attributed to the difference of the hybridization, resulted from the influence of the LDOS. Since $\mu=0$ here, the LDOS of the TLG around the Fermi level will play the major role in the hybridization. For example, in the r-TLG-$\textrm{A}_1$ case [ Fig. \ref{phase1} (f)], the LDOS has a peak at the Fermi level, which corresponds to a giant hybridization and thus greatly suppresses the magnetic phase. In contrast, the LDOS of $\textrm{B}_1$ site of r-TLG does not have a peak at $\mu=0$ ([see Fig. \ref{phase1} (h)], so that the magnetic region is larger.  The discussion above means that the adatom on the $A_1$ site of r-TLG prefers to be non-magnetic, compared with other cases.
We also notice that the phase diagram is symmetric around y=0. There are two reasons. One is that the bands for both b-TLG and r-TLG have particle-hole symmetry since we only consider the NN hopping. The other is that it is the undoped case with  $\mu=0$.

Then, we discuss the influence of the external electric field. For the case of b-TLG-$\textrm{B}_1$ [see Fig. \ref{phase1} (c)], the bias voltage $V_g$ will induce a LDOS peak near the Fermi level [see Fig. \ref{phase1} (d)], because of the electric field induced band overlap as shown in Fig. \ref{bands} (b). It will considerably enhance the hybridization near the Fermi level, so that the magnetic phase is suppressed [see Fig. \ref{phase1} (c)]. Meanwhile, due to the bias voltage, the LDOS is no longer symmetric around the Fermi level, and thus the phase diagram becomes asymmetric as well, as shown in Fig. \ref{phase1} (a) and (c).  As a comparison, we see that the modification of the LDOS (i.e. the hybridization) of the b-TLG-$\textrm{A}_1$ is not so obvious, so that the magnetic phase is not drastically changed, as given in Fig. \ref{phase1} (a).

\begin{figure}[t]
\centering
\includegraphics[width=8.5cm]{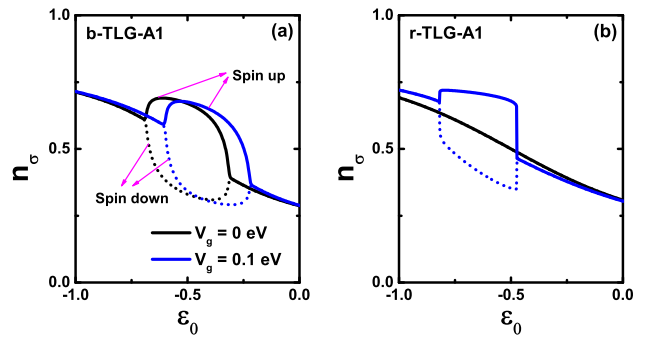}
\caption{(Color online) Spin resolved electron's occupation of the impurity as a function of the energy level with different $V_g$. $U = 1 \; \textrm{eV}$, $\mu = 0 \; \textrm{eV}$, $V = 3.2 \; \textrm{eV}$.
}
\label{manipulation}
\end{figure}

The r-TLG has a completely different behaviour, because that the bias voltage will induce an energy gap near the Fermi level [see Fig. \ref{bands} (a)]. For the case of r-TLG-$\textrm{A}_1$, we can see that the magnetic phase is enlarged  drastically by the applied bias voltage [see Fig. \ref{phase1} (e)]. This is because that, near $\mu=0$, the LDOS is changed from a peak ($V_g$=0) to nearly zero (finite $V_g$), where the hybridization has been greatly suppressed.  The nearly zero LDOS implies that the impurity level is decoupled from the TLG, which can be effectively considered as a free impurity level. So, very small $U$ can magnetize the impurity and the magnetic phase is enlarged.
The in-gap state in Fig. \ref{ingapmagnetic} actually corresponds to one point in the phase diagram of Fig. \ref{phase1} (e), where $x \approx 5$ and $y \approx 0.5$.

When the adatom is on the $\textrm{B}_1$ site, the undoped case of r-TLG is special. As we mentioned above, we can still change the impurity level into an in-gap state when the adatom is on $B_1$ site. But in Fig. \ref{phase1} (g), we see that  the change of the magnetic phase induced by electric field is not as obvious as the $A_1$ site case. The reason is that, for the undoped r-TLG, the effective hybridization here, i.e. LDOS at $B_1$ site, is  nearly zero occasionally
 [see Fig. \ref{phase1} (h)], so that the variation of the hybridization is tiny when the in-gap state appears. But as long as $\mu \neq 0$, as we will show later, the magnetic phase of the adatom on the $B_1$ site of r-TLG can be drastically changed by the electric field.

Note that $\varepsilon_d$ is the impurity level shifted by the bias voltage. Since the adatom is always on the top layer, the $\varepsilon_d$ of the positive and negative bias voltage are different. Thus, the magnetic phases for the positive and negative $V_g$  are not the same, as shown in Fig. \ref{phase1}.

The phenomena above are useful to manipulate the adatom  magnetic moment on TLG. For example, when a bias voltage is applied,
 it should be easier to realize the on/off transition of the magnetic moment  in the case of r-TLG-$\textrm{A}_1$ than in the b-TLG-$\textrm{A}_1$ case.  We give a concrete example in Fig. \ref{manipulation}, where the occupation of the impurity level is plotted for spin up and down, respectively. In Fig. \ref{manipulation} (a), we see that, for the case of b-TLG-$\textrm{A}_1$, only when the impurity level $\varepsilon_0$ is in a very small  region, a bias voltage can realize the on/off transition of the local magnetic moment.  Namely, when the $\varepsilon_0$ is near $-0.6$ eV, the local moment (i.e. $n_{\uparrow}-n_{\downarrow}$) is finite for $V_g=0$ but zero if a bias voltage ($V_g=0.1$ eV) is applied.
   In contrast, for the r-TLG-$\textrm{A}_1$ case [see Fig. \ref{manipulation} (b)], the magnetic moment can be switched by a bias voltage within a much wider allowed energy region of the impurity level, i.e. between $-0.8$ eV and $-0.5$ eV.

\begin{figure}[t!]
\centering
\includegraphics[width=8.5cm]{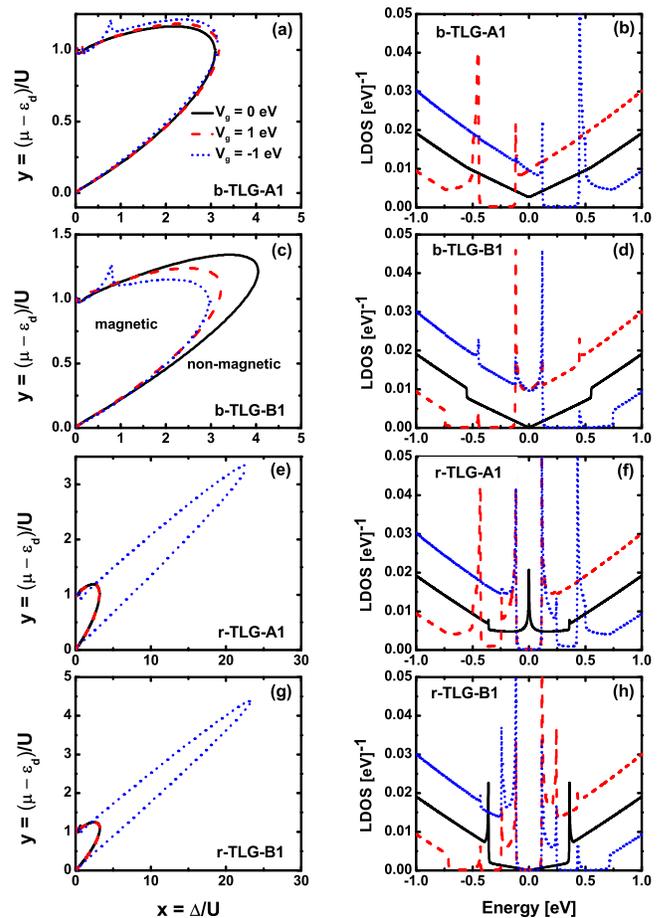}
\caption{(Color online) (a),(c),(e),(g): The magnetic phase boundaries for an impurity on TLG with $\varepsilon_{0} = - 0.5 \; \textrm{eV}$. (b),(d),(f),(h): The corresponding LDOS of TLG at the adsorption site.
 $V = 0.5 \; \textrm{eV}$.
}
\label{phase2}
\end{figure}

All the discussions above are about the undoped TLG with $\mu=0$. In principle,  the doping and bias voltage can be controlled separately in the dual-gate configuration in experiment. It should be noted that, in TLG, the effects of doping and applied bias voltage are different, where doping tunes the chemical potential but a bias voltage changes the band structure.  Thus, we then give the  the phase boundary of the adatom with fixed $\varepsilon_0$ in Fig. \ref{phase2}, which means that the impurity level is given and the chemical potential (doping)  is tunable.

First, we see that the magnetic phase is no longer symmetric around $y=0.5$, which results from the fact that the particle-hole symmetry is absent when $\mu$ is nonzero. This is similar as the cases of monolayer and bilayer graphene in former literatures. Some interesting results appear when a bias voltage is applied. The influence of the electric field on the magnetic phase is  very significant for the r-TLG, while it is tiny for b-TLG.
And, the impurity magnetic moment in r-TLG depends strongly on the sign of the applied bias $V_g$, i.e. the direction of the applied electric field.
 We plot the magnetic phases for $V_g = \pm 1$ eV.  As demonstrated in Fig. \ref{phase2} (e) and (g), a negative bias voltage $V_g=-1$ eV can greatly enlarge the magnetic phase, but a positive bias voltage can not. This is because that, since $\varepsilon_0 = -0.5$ eV here, as discussed in last section, a negative bias voltage can make the impurity level into an in-gap state, which prefers to be magnetic even with tiny on-site Coulomb interaction.  Thus, the modification of  magnetic phase is so drastic.
But for a positive bias, the in-gap state can not be achieved, so that the change of the magnetic phase is small.
 Similarly,  the phase diagram of b-TLG is slightly changed by the bias voltage, as given in Fig. \ref{phase2} (a) and (c), due to the absence of the in-gap state.  Different from the former case, because that $\mu$ is a variable here, we do not find an intuitive relation between the LDOS and the magnetic phase, though the LDOS is also given for comparison.

\section{summary}\label{summary}
In summary, we theoretically investigate the adatom on the top-site of  TLG by solving the  the Anderson impurity model via self-consistent mean field method. The influences of the stacking order, nonequivalent adsorption sites, and external electric field are carefully considered.

 Our main findings are: \textbf{(1)} Due to the divergent DOS at the Fermi level, the adatom on the $\textrm{A}_1$ site of r-TLG can have a Fano-shaped impurity spectral density, which may be a general property for the impurity level near the metal DOS singularity. The mechanism is similar as the case of Kondo peak induced Fano resonance, which has been intensively studied in the last two decades. \textbf{(2)} For the top-site adatom on r-TLG, the impurity level can be tuned into an in-gap state by an external electric field,  which depends strongly on the polarity of the applied bias. The in-gap state  greatly modifies the local moment formation of the adatom. \textbf{(3)} Except for adjusting the chemical potential, we demonstrate that controlling the hybridization between the impurity level and the conducting electrons via a bias voltage is another effective way to manipulate the magnetic moment of the adatom on TLG.   We  systematically calculate the magnetic phase diagram of the adatom on undoped TLG in the presence of an external electric field for various cases.
 For undoped r-TLG-$\textrm{A}_1$, the divergent DOS at the Fermi level will greatly enhance the hybridization, so that the adatom is unlikely to be magnetic.
  But if a bias voltage is applied, a gap is opened, hybridization is decreased and thus, the adatom can be tuned to be magnetic. Furthermore, if the impurity level is tuned into an in-gap state by a proper bias, tiny on-site Coulomb interaction can make the adatom to be magnetic, which greatly enlarges the magnetic phase in the phase diagram. The situations of the b-TLG are in  contradiction to that of the r-TLG, because that electric field will induce a band overlap and increase the hybridization. The typical example is the case of b-TLG-$\textrm{B}_1$,  where the electric field enhanced hybridization will obviously restrain the magnetic phase. At last, we  discuss the influence of doping on the magnetic phase diagram for the TLG.
  We show that the impurity magnetic phase of r-TLG strongly depends on the direction of the electric field, because of the electric-field-induced in-gap state.

We would like to compare the situations of the monolayer\cite{prl2008,jafarijpcm,uchoanjp} and bilayer graphene\cite{ding2009,bilayer2011,bilayer2014,aabilayer} with the case of TLG we studied here. The DOS of the undoped monolayer graphene is zero at the Fermi level , so that the effective hybridization is nearly zero and the adatom prefers to be magnetic. Meanwhile, in monolayer graphene, the effect of  electric field is merely to change the chemical potential (doping). As for the bilayer graphene, it is in some sense similar as r-TLG, because that a bias voltage (or an electric field) can open a gap. We think that the electric-field-induced in-gap state will also appear in the bilayer case.   But, for the undoped bilayer graphene, the DOS at the Fermi level is a constant, not divergent as r-TLG. Thus, it should be easier to get a magnetic adatom in bilayer  than in r-TLG. And there should be no Fano shaped impurity spectral in bilayer graphene.  Meanwhile, only in the b-TLG, the electric field can enhance the hybridization and suppress the local magnetic moment.  We argue that the influence of the electric field on the adatom in TLG is more significant than that in the bilayer graphene.

Finally, let us discuss the related experiments. The most suitable experimental system to test our theoretical results is the hydrogen atom absorbed on TLG, which can be exactly described by the single level Anderson impurity model. In experiment,  the hydrogen atom favors the top site of graphene, and the magnetic moment of hydrogen atom on graphene has been confirmed recently via STM measurement\cite{science2016}. We assume that, on the TLG, the hydrogen atom should also favor the top site, not the hollow and bridge sites.  The magnetism of hydrogen adatom can  be readily detected by the STM.  Furthermore,   the DOS peak of the undoped  of r-TLG near the Fermi level has been observed recently also by the STM\cite{xiao2015}. The bias voltage induced gap has also been confirmed by various experimental techniques, including STM\cite{stmtrilayergap}, transport\cite{abctransport} and optical measurements\cite{heinz2011}. So, in principle, there is no fundamental difficulties to test our results in experiment. We believe that our work gives a comprehensive understanding about the magnetic moment of the top-site adatom on TLG, and hope that it can stimulate further theoretical and experimental interest in this intriguing issue.

\begin{acknowledgments}
J.H.G is supported by the National Natural Science Foundation of China (Grants  No. 11534001, No. 11274129,). J.T.Lu is supported by the National Natural Science Foundation of China (Grant Nos. 61371015 and 11304107).
\end{acknowledgments}

\bibliography{gaotrilayerbib}

\begin{thebibliography}{55}%
\makeatletter
\providecommand \@ifxundefined [1]{%
 \@ifx{#1\undefined}
}%
\providecommand \@ifnum [1]{%
 \ifnum #1\expandafter \@firstoftwo
 \else \expandafter \@secondoftwo
 \fi
}%
\providecommand \@ifx [1]{%
 \ifx #1\expandafter \@firstoftwo
 \else \expandafter \@secondoftwo
 \fi
}%
\providecommand \natexlab [1]{#1}%
\providecommand \enquote  [1]{``#1''}%
\providecommand \bibnamefont  [1]{#1}%
\providecommand \bibfnamefont [1]{#1}%
\providecommand \citenamefont [1]{#1}%
\providecommand \href@noop [0]{\@secondoftwo}%
\providecommand \href [0]{\begingroup \@sanitize@url \@href}%
\providecommand \@href[1]{\@@startlink{#1}\@@href}%
\providecommand \@@href[1]{\endgroup#1\@@endlink}%
\providecommand \@sanitize@url [0]{\catcode `\\12\catcode `\$12\catcode
  `\&12\catcode `\#12\catcode `\^12\catcode `\_12\catcode `\%12\relax}%
\providecommand \@@startlink[1]{}%
\providecommand \@@endlink[0]{}%
\providecommand \url  [0]{\begingroup\@sanitize@url \@url }%
\providecommand \@url [1]{\endgroup\@href {#1}{\urlprefix }}%
\providecommand \urlprefix  [0]{URL }%
\providecommand \Eprint [0]{\href }%
\providecommand \doibase [0]{http://dx.doi.org/}%
\providecommand \selectlanguage [0]{\@gobble}%
\providecommand \bibinfo  [0]{\@secondoftwo}%
\providecommand \bibfield  [0]{\@secondoftwo}%
\providecommand \translation [1]{[#1]}%
\providecommand \BibitemOpen [0]{}%
\providecommand \bibitemStop [0]{}%
\providecommand \bibitemNoStop [0]{.\EOS\space}%
\providecommand \EOS [0]{\spacefactor3000\relax}%
\providecommand \BibitemShut  [1]{\csname bibitem#1\endcsname}%
\let\auto@bib@innerbib\@empty
%</preamble>
\bibitem [{\citenamefont {Wehling}\ \emph {et~al.}(2009)\citenamefont
  {Wehling}, \citenamefont {Katsnelson},\ and\ \citenamefont
  {Lichtenstein}}]{chemreview2009}%
  \BibitemOpen
  \bibfield  {author} {\bibinfo {author} {\bibfnamefont {T.~O.}\ \bibnamefont
  {Wehling}}, \bibinfo {author} {\bibfnamefont {M.~I.}\ \bibnamefont
  {Katsnelson}}, \ and\ \bibinfo {author} {\bibfnamefont {A.~I.}\ \bibnamefont
  {Lichtenstein}},\ }\href {\doibase 10.1016/j.cplett.2009.06.005} {\bibfield
  {journal} {\bibinfo  {journal} {Chem. Phys. Lett.}\ }\textbf {\bibinfo
  {volume} {476}},\ \bibinfo {pages} {125} (\bibinfo {year}
  {2009})}\BibitemShut {NoStop}%
\bibitem [{\citenamefont {Katoch}(2015)}]{adatomreview}%
  \BibitemOpen
  \bibfield  {author} {\bibinfo {author} {\bibfnamefont {J.}~\bibnamefont
  {Katoch}},\ }\href {\doibase
  http://dx.doi.org/10.1016/j.synthmet.2015.07.017} {\bibfield  {journal}
  {\bibinfo  {journal} {Synth. Met.}\ }\textbf {\bibinfo {volume} {210, Part
  A}},\ \bibinfo {pages} {68 } (\bibinfo {year} {2015})}\BibitemShut {NoStop}%
\bibitem [{\citenamefont {Chen}\ \emph {et~al.}(2008)\citenamefont {Chen},
  \citenamefont {Jang}, \citenamefont {Adam}, \citenamefont {Fuhrer},
  \citenamefont {Williams},\ and\ \citenamefont {Ishigami}}]{chenjh2008}%
  \BibitemOpen
  \bibfield  {author} {\bibinfo {author} {\bibfnamefont {J.-H.}\ \bibnamefont
  {Chen}}, \bibinfo {author} {\bibfnamefont {C.}~\bibnamefont {Jang}}, \bibinfo
  {author} {\bibfnamefont {S.}~\bibnamefont {Adam}}, \bibinfo {author}
  {\bibfnamefont {M.~S.}\ \bibnamefont {Fuhrer}}, \bibinfo {author}
  {\bibfnamefont {E.~D.}\ \bibnamefont {Williams}}, \ and\ \bibinfo {author}
  {\bibfnamefont {M.}~\bibnamefont {Ishigami}},\ }\href {\doibase
  10.1038/nphys935} {\bibfield  {journal} {\bibinfo  {journal} {Nat. Phys.}\
  }\textbf {\bibinfo {volume} {4}},\ \bibinfo {pages} {377} (\bibinfo {year}
  {2008})}\BibitemShut {NoStop}%
\bibitem [{\citenamefont {Katoch}\ and\ \citenamefont
  {Ishigami}(2012)}]{solid2012}%
  \BibitemOpen
  \bibfield  {author} {\bibinfo {author} {\bibfnamefont {J.}~\bibnamefont
  {Katoch}}\ and\ \bibinfo {author} {\bibfnamefont {M.}~\bibnamefont
  {Ishigami}},\ }\href {\doibase http://dx.doi.org/10.1016/j.ssc.2011.11.003}
  {\bibfield  {journal} {\bibinfo  {journal} {Solid State Commun.}\ }\textbf
  {\bibinfo {volume} {152}},\ \bibinfo {pages} {60 } (\bibinfo {year}
  {2012})}\BibitemShut {NoStop}%
\bibitem [{\citenamefont {Yan}\ and\ \citenamefont {Fuhrer}(2011)}]{prl2011}%
  \BibitemOpen
  \bibfield  {author} {\bibinfo {author} {\bibfnamefont {J.}~\bibnamefont
  {Yan}}\ and\ \bibinfo {author} {\bibfnamefont {M.~S.}\ \bibnamefont
  {Fuhrer}},\ }\href {\doibase 10.1103/PhysRevLett.107.206601} {\bibfield
  {journal} {\bibinfo  {journal} {Phys. Rev. Lett.}\ }\textbf {\bibinfo
  {volume} {107}},\ \bibinfo {pages} {206601} (\bibinfo {year}
  {2011})}\BibitemShut {NoStop}%
\bibitem [{\citenamefont {Wehling}\ \emph {et~al.}(2010)\citenamefont
  {Wehling}, \citenamefont {Yuan}, \citenamefont {Lichtenstein}, \citenamefont
  {Geim},\ and\ \citenamefont {Katsnelson}}]{katsnelson2010}%
  \BibitemOpen
  \bibfield  {author} {\bibinfo {author} {\bibfnamefont {T.~O.}\ \bibnamefont
  {Wehling}}, \bibinfo {author} {\bibfnamefont {S.}~\bibnamefont {Yuan}},
  \bibinfo {author} {\bibfnamefont {A.~I.}\ \bibnamefont {Lichtenstein}},
  \bibinfo {author} {\bibfnamefont {A.~K.}\ \bibnamefont {Geim}}, \ and\
  \bibinfo {author} {\bibfnamefont {M.~I.}\ \bibnamefont {Katsnelson}},\ }\href
  {\doibase 10.1103/PhysRevLett.105.056802} {\bibfield  {journal} {\bibinfo
  {journal} {Phys. Rev. Lett.}\ }\textbf {\bibinfo {volume} {105}},\ \bibinfo
  {pages} {056802} (\bibinfo {year} {2010})}\BibitemShut {NoStop}%
\bibitem [{\citenamefont {Weeks}\ \emph {et~al.}(2011)\citenamefont {Weeks},
  \citenamefont {Hu}, \citenamefont {Alicea}, \citenamefont {Franz},\ and\
  \citenamefont {Wu}}]{wu2011}%
  \BibitemOpen
  \bibfield  {author} {\bibinfo {author} {\bibfnamefont {C.}~\bibnamefont
  {Weeks}}, \bibinfo {author} {\bibfnamefont {J.}~\bibnamefont {Hu}}, \bibinfo
  {author} {\bibfnamefont {J.}~\bibnamefont {Alicea}}, \bibinfo {author}
  {\bibfnamefont {M.}~\bibnamefont {Franz}}, \ and\ \bibinfo {author}
  {\bibfnamefont {R.}~\bibnamefont {Wu}},\ }\href {\doibase
  10.1103/PhysRevX.1.021001} {\bibfield  {journal} {\bibinfo  {journal} {Phys.
  Rev. X}\ }\textbf {\bibinfo {volume} {1}},\ \bibinfo {pages} {021001}
  (\bibinfo {year} {2011})}\BibitemShut {NoStop}%
\bibitem [{\citenamefont {Hu}\ \emph {et~al.}(2012)\citenamefont {Hu},
  \citenamefont {Alicea}, \citenamefont {Wu},\ and\ \citenamefont
  {Franz}}]{hujun2012}%
  \BibitemOpen
  \bibfield  {author} {\bibinfo {author} {\bibfnamefont {J.}~\bibnamefont
  {Hu}}, \bibinfo {author} {\bibfnamefont {J.}~\bibnamefont {Alicea}}, \bibinfo
  {author} {\bibfnamefont {R.}~\bibnamefont {Wu}}, \ and\ \bibinfo {author}
  {\bibfnamefont {M.}~\bibnamefont {Franz}},\ }\href {\doibase
  10.1103/PhysRevLett.109.266801} {\bibfield  {journal} {\bibinfo  {journal}
  {Phys. Rev. Lett.}\ }\textbf {\bibinfo {volume} {109}},\ \bibinfo {pages}
  {266801} (\bibinfo {year} {2012})}\BibitemShut {NoStop}%
\bibitem [{\citenamefont {Jiang}\ \emph {et~al.}(2012)\citenamefont {Jiang},
  \citenamefont {Qiao}, \citenamefont {Liu}, \citenamefont {Shi},\ and\
  \citenamefont {Niu}}]{jianghua2012}%
  \BibitemOpen
  \bibfield  {author} {\bibinfo {author} {\bibfnamefont {H.}~\bibnamefont
  {Jiang}}, \bibinfo {author} {\bibfnamefont {Z.}~\bibnamefont {Qiao}},
  \bibinfo {author} {\bibfnamefont {H.}~\bibnamefont {Liu}}, \bibinfo {author}
  {\bibfnamefont {J.}~\bibnamefont {Shi}}, \ and\ \bibinfo {author}
  {\bibfnamefont {Q.}~\bibnamefont {Niu}},\ }\href {\doibase
  10.1103/PhysRevLett.109.116803} {\bibfield  {journal} {\bibinfo  {journal}
  {Phys. Rev. Lett.}\ }\textbf {\bibinfo {volume} {109}},\ \bibinfo {pages}
  {116803} (\bibinfo {year} {2012})}\BibitemShut {NoStop}%
\bibitem [{\citenamefont {Gmitra}\ \emph {et~al.}(2013)\citenamefont {Gmitra},
  \citenamefont {Kochan},\ and\ \citenamefont {Fabian}}]{hydrogenSO}%
  \BibitemOpen
  \bibfield  {author} {\bibinfo {author} {\bibfnamefont {M.}~\bibnamefont
  {Gmitra}}, \bibinfo {author} {\bibfnamefont {D.}~\bibnamefont {Kochan}}, \
  and\ \bibinfo {author} {\bibfnamefont {J.}~\bibnamefont {Fabian}},\ }\href
  {\doibase 10.1103/PhysRevLett.110.246602} {\bibfield  {journal} {\bibinfo
  {journal} {Phys. Rev. Lett.}\ }\textbf {\bibinfo {volume} {110}},\ \bibinfo
  {pages} {246602} (\bibinfo {year} {2013})}\BibitemShut {NoStop}%
\bibitem [{\citenamefont {Liu}\ \emph {et~al.}(2015)\citenamefont {Liu},
  \citenamefont {Zhu},\ and\ \citenamefont {Zheng}}]{zheng2015}%
  \BibitemOpen
  \bibfield  {author} {\bibinfo {author} {\bibfnamefont {Z.}~\bibnamefont
  {Liu}}, \bibinfo {author} {\bibfnamefont {M.}~\bibnamefont {Zhu}}, \ and\
  \bibinfo {author} {\bibfnamefont {Y.}~\bibnamefont {Zheng}},\ }\href
  {\doibase 10.1103/PhysRevB.92.245438} {\bibfield  {journal} {\bibinfo
  {journal} {Phys. Rev. B}\ }\textbf {\bibinfo {volume} {92}},\ \bibinfo
  {pages} {245438} (\bibinfo {year} {2015})}\BibitemShut {NoStop}%
\bibitem [{\citenamefont {Jia}\ \emph {et~al.}(2015)\citenamefont {Jia},
  \citenamefont {Yan}, \citenamefont {Niu}, \citenamefont {Han}, \citenamefont
  {Zhu}, \citenamefont {Yu},\ and\ \citenamefont {Wu}}]{wuxiaosong2015}%
  \BibitemOpen
  \bibfield  {author} {\bibinfo {author} {\bibfnamefont {Z.}~\bibnamefont
  {Jia}}, \bibinfo {author} {\bibfnamefont {B.}~\bibnamefont {Yan}}, \bibinfo
  {author} {\bibfnamefont {J.}~\bibnamefont {Niu}}, \bibinfo {author}
  {\bibfnamefont {Q.}~\bibnamefont {Han}}, \bibinfo {author} {\bibfnamefont
  {R.}~\bibnamefont {Zhu}}, \bibinfo {author} {\bibfnamefont {D.}~\bibnamefont
  {Yu}}, \ and\ \bibinfo {author} {\bibfnamefont {X.}~\bibnamefont {Wu}},\
  }\href {\doibase 10.1103/PhysRevB.91.085411} {\bibfield  {journal} {\bibinfo
  {journal} {Phys. Rev. B}\ }\textbf {\bibinfo {volume} {91}},\ \bibinfo
  {pages} {085411} (\bibinfo {year} {2015})}\BibitemShut {NoStop}%
\bibitem [{\citenamefont {Chandni}\ \emph {et~al.}(2015)\citenamefont
  {Chandni}, \citenamefont {Henriksen},\ and\ \citenamefont
  {Eisenstein}}]{indium2015}%
  \BibitemOpen
  \bibfield  {author} {\bibinfo {author} {\bibfnamefont {U.}~\bibnamefont
  {Chandni}}, \bibinfo {author} {\bibfnamefont {E.~A.}\ \bibnamefont
  {Henriksen}}, \ and\ \bibinfo {author} {\bibfnamefont {J.~P.}\ \bibnamefont
  {Eisenstein}},\ }\href {\doibase 10.1103/PhysRevB.91.245402} {\bibfield
  {journal} {\bibinfo  {journal} {Phys. Rev. B}\ }\textbf {\bibinfo {volume}
  {91}},\ \bibinfo {pages} {245402} (\bibinfo {year} {2015})}\BibitemShut
  {NoStop}%
\bibitem [{\citenamefont {Uchoa}\ \emph {et~al.}(2008)\citenamefont {Uchoa},
  \citenamefont {Kotov}, \citenamefont {Peres},\ and\ \citenamefont
  {Castro~Neto}}]{prl2008}%
  \BibitemOpen
  \bibfield  {author} {\bibinfo {author} {\bibfnamefont {B.}~\bibnamefont
  {Uchoa}}, \bibinfo {author} {\bibfnamefont {V.~N.}\ \bibnamefont {Kotov}},
  \bibinfo {author} {\bibfnamefont {N.~M.~R.}\ \bibnamefont {Peres}}, \ and\
  \bibinfo {author} {\bibfnamefont {A.~H.}\ \bibnamefont {Castro~Neto}},\
  }\href {\doibase 10.1103/PhysRevLett.101.026805} {\bibfield  {journal}
  {\bibinfo  {journal} {Phys. Rev. Lett.}\ }\textbf {\bibinfo {volume} {101}},\
  \bibinfo {pages} {026805} (\bibinfo {year} {2008})}\BibitemShut {NoStop}%
\bibitem [{\citenamefont {Uchoa}\ \emph {et~al.}(2014)\citenamefont {Uchoa},
  \citenamefont {Yang}, \citenamefont {Tsai}, \citenamefont {Peres},\ and\
  \citenamefont {Neto}}]{uchoanjp}%
  \BibitemOpen
  \bibfield  {author} {\bibinfo {author} {\bibfnamefont {B.}~\bibnamefont
  {Uchoa}}, \bibinfo {author} {\bibfnamefont {L.}~\bibnamefont {Yang}},
  \bibinfo {author} {\bibfnamefont {S.-W.}\ \bibnamefont {Tsai}}, \bibinfo
  {author} {\bibfnamefont {N.~M.~R.}\ \bibnamefont {Peres}}, \ and\ \bibinfo
  {author} {\bibfnamefont {A.~H.~C.}\ \bibnamefont {Neto}},\ }\href
  {http://stacks.iop.org/1367-2630/16/i=1/a=013045} {\bibfield  {journal}
  {\bibinfo  {journal} {New J. Phys.}\ }\textbf {\bibinfo {volume} {16}},\
  \bibinfo {pages} {013045} (\bibinfo {year} {2014})}\BibitemShut {NoStop}%
\bibitem [{\citenamefont {McCreary}\ \emph {et~al.}(2012)\citenamefont
  {McCreary}, \citenamefont {Swartz}, \citenamefont {Han}, \citenamefont
  {Fabian},\ and\ \citenamefont {Kawakami}}]{hydrogentransport}%
  \BibitemOpen
  \bibfield  {author} {\bibinfo {author} {\bibfnamefont {K.~M.}\ \bibnamefont
  {McCreary}}, \bibinfo {author} {\bibfnamefont {A.~G.}\ \bibnamefont
  {Swartz}}, \bibinfo {author} {\bibfnamefont {W.}~\bibnamefont {Han}},
  \bibinfo {author} {\bibfnamefont {J.}~\bibnamefont {Fabian}}, \ and\ \bibinfo
  {author} {\bibfnamefont {R.~K.}\ \bibnamefont {Kawakami}},\ }\href {\doibase
  10.1103/PhysRevLett.109.186604} {\bibfield  {journal} {\bibinfo  {journal}
  {Phys. Rev. Lett.}\ }\textbf {\bibinfo {volume} {109}},\ \bibinfo {pages}
  {186604} (\bibinfo {year} {2012})}\BibitemShut {NoStop}%
\bibitem [{\citenamefont {Mashkoori}\ and\ \citenamefont
  {Jafari}(2015)}]{jafarijpcm}%
  \BibitemOpen
  \bibfield  {author} {\bibinfo {author} {\bibfnamefont {M.}~\bibnamefont
  {Mashkoori}}\ and\ \bibinfo {author} {\bibfnamefont {S.~A.}\ \bibnamefont
  {Jafari}},\ }\href {http://stacks.iop.org/0953-8984/27/i=15/a=156001}
  {\bibfield  {journal} {\bibinfo  {journal} {J. Phys.: Condens. Matter}\
  }\textbf {\bibinfo {volume} {27}},\ \bibinfo {pages} {156001} (\bibinfo
  {year} {2015})}\BibitemShut {NoStop}%
\bibitem [{\citenamefont {Gonz{\'a}lez-Herrero}\ \emph
  {et~al.}(2016)\citenamefont {Gonz{\'a}lez-Herrero}, \citenamefont
  {G{\'o}mez-Rodr{\'\i}guez}, \citenamefont {Mallet}, \citenamefont {Moaied},
  \citenamefont {Palacios}, \citenamefont {Salgado}, \citenamefont {Ugeda},
  \citenamefont {Veuillen}, \citenamefont {Yndurain},\ and\ \citenamefont
  {Brihuega}}]{science2016}%
  \BibitemOpen
  \bibfield  {author} {\bibinfo {author} {\bibfnamefont {H.}~\bibnamefont
  {Gonz{\'a}lez-Herrero}}, \bibinfo {author} {\bibfnamefont {J.~M.}\
  \bibnamefont {G{\'o}mez-Rodr{\'\i}guez}}, \bibinfo {author} {\bibfnamefont
  {P.}~\bibnamefont {Mallet}}, \bibinfo {author} {\bibfnamefont
  {M.}~\bibnamefont {Moaied}}, \bibinfo {author} {\bibfnamefont {J.~J.}\
  \bibnamefont {Palacios}}, \bibinfo {author} {\bibfnamefont {C.}~\bibnamefont
  {Salgado}}, \bibinfo {author} {\bibfnamefont {M.~M.}\ \bibnamefont {Ugeda}},
  \bibinfo {author} {\bibfnamefont {J.-Y.}\ \bibnamefont {Veuillen}}, \bibinfo
  {author} {\bibfnamefont {F.}~\bibnamefont {Yndurain}}, \ and\ \bibinfo
  {author} {\bibfnamefont {I.}~\bibnamefont {Brihuega}},\ }\href {\doibase
  10.1126/science.aad8038} {\bibfield  {journal} {\bibinfo  {journal}
  {Science}\ }\textbf {\bibinfo {volume} {352}},\ \bibinfo {pages} {437}
  (\bibinfo {year} {2016})}\BibitemShut {NoStop}%
\bibitem [{\citenamefont {Fritz}\ and\ \citenamefont
  {Vojta}(2013)}]{kondoreview}%
  \BibitemOpen
  \bibfield  {author} {\bibinfo {author} {\bibfnamefont {L.}~\bibnamefont
  {Fritz}}\ and\ \bibinfo {author} {\bibfnamefont {M.}~\bibnamefont {Vojta}},\
  }\href {http://stacks.iop.org/0034-4885/76/i=3/a=032501} {\bibfield
  {journal} {\bibinfo  {journal} {Rep. Prog. Phys.}\ }\textbf {\bibinfo
  {volume} {76}},\ \bibinfo {pages} {032501} (\bibinfo {year}
  {2013})}\BibitemShut {NoStop}%
\bibitem [{\citenamefont {Vojta}\ \emph {et~al.}(2010)\citenamefont {Vojta},
  \citenamefont {Fritz},\ and\ \citenamefont {Bulla}}]{bulla2010}%
  \BibitemOpen
  \bibfield  {author} {\bibinfo {author} {\bibfnamefont {M.}~\bibnamefont
  {Vojta}}, \bibinfo {author} {\bibfnamefont {L.}~\bibnamefont {Fritz}}, \ and\
  \bibinfo {author} {\bibfnamefont {R.}~\bibnamefont {Bulla}},\ }\href
  {\doibase 10.1209/0295-5075/90/27006} {\bibfield  {journal} {\bibinfo
  {journal} {Europhys. Lett.}\ }\textbf {\bibinfo {volume} {90}},\ \bibinfo
  {pages} {27006} (\bibinfo {year} {2010})}\BibitemShut {NoStop}%
\bibitem [{\citenamefont {Li}\ \emph {et~al.}(2013)\citenamefont {Li},
  \citenamefont {Ni}, \citenamefont {Zhong}, \citenamefont {Fang},\ and\
  \citenamefont {Luo}}]{lilinnjp}%
  \BibitemOpen
  \bibfield  {author} {\bibinfo {author} {\bibfnamefont {L.}~\bibnamefont
  {Li}}, \bibinfo {author} {\bibfnamefont {Y.-Y.}\ \bibnamefont {Ni}}, \bibinfo
  {author} {\bibfnamefont {Y.}~\bibnamefont {Zhong}}, \bibinfo {author}
  {\bibfnamefont {T.-F.}\ \bibnamefont {Fang}}, \ and\ \bibinfo {author}
  {\bibfnamefont {H.-G.}\ \bibnamefont {Luo}},\ }\href
  {http://stacks.iop.org/1367-2630/15/i=5/a=053018} {\bibfield  {journal}
  {\bibinfo  {journal} {New J. Phys.}\ }\textbf {\bibinfo {volume} {15}},\
  \bibinfo {pages} {053018} (\bibinfo {year} {2013})}\BibitemShut {NoStop}%
\bibitem [{\citenamefont {Zhuang}\ \emph {et~al.}(2009)\citenamefont {Zhuang},
  \citenamefont {feng Sun},\ and\ \citenamefont {Xie}}]{zhuanghb2009}%
  \BibitemOpen
  \bibfield  {author} {\bibinfo {author} {\bibfnamefont {H.-B.}\ \bibnamefont
  {Zhuang}}, \bibinfo {author} {\bibfnamefont {Q.}~\bibnamefont {feng Sun}}, \
  and\ \bibinfo {author} {\bibfnamefont {X.~C.}\ \bibnamefont {Xie}},\ }\href
  {http://stacks.iop.org/0295-5075/86/i=5/a=58004} {\bibfield  {journal}
  {\bibinfo  {journal} {Europhys. Lett.}\ }\textbf {\bibinfo {volume} {86}},\
  \bibinfo {pages} {58004} (\bibinfo {year} {2009})}\BibitemShut {NoStop}%
\bibitem [{\citenamefont {Uchoa}\ \emph {et~al.}(2011)\citenamefont {Uchoa},
  \citenamefont {Rappoport},\ and\ \citenamefont
  {Castro~Neto}}]{uchoakondo2011}%
  \BibitemOpen
  \bibfield  {author} {\bibinfo {author} {\bibfnamefont {B.}~\bibnamefont
  {Uchoa}}, \bibinfo {author} {\bibfnamefont {T.~G.}\ \bibnamefont
  {Rappoport}}, \ and\ \bibinfo {author} {\bibfnamefont {A.~H.}\ \bibnamefont
  {Castro~Neto}},\ }\href {\doibase 10.1103/PhysRevLett.106.016801} {\bibfield
  {journal} {\bibinfo  {journal} {Phys. Rev. Lett.}\ }\textbf {\bibinfo
  {volume} {106}},\ \bibinfo {pages} {016801} (\bibinfo {year}
  {2011})}\BibitemShut {NoStop}%
\bibitem [{\citenamefont {Anderson}(1961)}]{anderson61}%
  \BibitemOpen
  \bibfield  {author} {\bibinfo {author} {\bibfnamefont {P.~W.}\ \bibnamefont
  {Anderson}},\ }\href {\doibase 10.1103/PhysRev.124.41} {\bibfield  {journal}
  {\bibinfo  {journal} {Phys. Rev.}\ }\textbf {\bibinfo {volume} {124}},\
  \bibinfo {pages} {41} (\bibinfo {year} {1961})}\BibitemShut {NoStop}%
\bibitem [{\citenamefont {Castro~Neto}\ \emph {et~al.}(2009)\citenamefont
  {Castro~Neto}, \citenamefont {Guinea}, \citenamefont {Peres}, \citenamefont
  {Novoselov},\ and\ \citenamefont {Geim}}]{rmpgraphene}%
  \BibitemOpen
  \bibfield  {author} {\bibinfo {author} {\bibfnamefont {A.~H.}\ \bibnamefont
  {Castro~Neto}}, \bibinfo {author} {\bibfnamefont {F.}~\bibnamefont {Guinea}},
  \bibinfo {author} {\bibfnamefont {N.~M.~R.}\ \bibnamefont {Peres}}, \bibinfo
  {author} {\bibfnamefont {K.~S.}\ \bibnamefont {Novoselov}}, \ and\ \bibinfo
  {author} {\bibfnamefont {A.~K.}\ \bibnamefont {Geim}},\ }\href {\doibase
  10.1103/RevModPhys.81.109} {\bibfield  {journal} {\bibinfo  {journal} {Rev.
  Mod. Phys.}\ }\textbf {\bibinfo {volume} {81}},\ \bibinfo {pages} {109}
  (\bibinfo {year} {2009})}\BibitemShut {NoStop}%
\bibitem [{\citenamefont {Guinea}\ \emph {et~al.}(2006)\citenamefont {Guinea},
  \citenamefont {Castro~Neto},\ and\ \citenamefont {Peres}}]{guinea2006}%
  \BibitemOpen
  \bibfield  {author} {\bibinfo {author} {\bibfnamefont {F.}~\bibnamefont
  {Guinea}}, \bibinfo {author} {\bibfnamefont {A.~H.}\ \bibnamefont
  {Castro~Neto}}, \ and\ \bibinfo {author} {\bibfnamefont {N.~M.~R.}\
  \bibnamefont {Peres}},\ }\href {\doibase 10.1103/PhysRevB.73.245426}
  {\bibfield  {journal} {\bibinfo  {journal} {Phys. Rev. B}\ }\textbf {\bibinfo
  {volume} {73}},\ \bibinfo {pages} {245426} (\bibinfo {year}
  {2006})}\BibitemShut {NoStop}%
\bibitem [{\citenamefont {Partoens}\ and\ \citenamefont
  {Peeters}(2006)}]{aba2006}%
  \BibitemOpen
  \bibfield  {author} {\bibinfo {author} {\bibfnamefont {B.}~\bibnamefont
  {Partoens}}\ and\ \bibinfo {author} {\bibfnamefont {F.~M.}\ \bibnamefont
  {Peeters}},\ }\href {\doibase 10.1103/PhysRevB.74.075404} {\bibfield
  {journal} {\bibinfo  {journal} {Phys. Rev. B}\ }\textbf {\bibinfo {volume}
  {74}},\ \bibinfo {pages} {075404} (\bibinfo {year} {2006})}\BibitemShut
  {NoStop}%
\bibitem [{\citenamefont {Aoki}\ and\ \citenamefont
  {Amawashi}(2007)}]{solidtrilayer}%
  \BibitemOpen
  \bibfield  {author} {\bibinfo {author} {\bibfnamefont {M.}~\bibnamefont
  {Aoki}}\ and\ \bibinfo {author} {\bibfnamefont {H.}~\bibnamefont
  {Amawashi}},\ }\href {\doibase http://dx.doi.org/10.1016/j.ssc.2007.02.013}
  {\bibfield  {journal} {\bibinfo  {journal} {Solid State Commun.}\ }\textbf
  {\bibinfo {volume} {142}},\ \bibinfo {pages} {123 } (\bibinfo {year}
  {2007})}\BibitemShut {NoStop}%
\bibitem [{\citenamefont {Koshino}\ and\ \citenamefont
  {McCann}(2009)}]{abakoshino}%
  \BibitemOpen
  \bibfield  {author} {\bibinfo {author} {\bibfnamefont {M.}~\bibnamefont
  {Koshino}}\ and\ \bibinfo {author} {\bibfnamefont {E.}~\bibnamefont
  {McCann}},\ }\href {\doibase 10.1103/PhysRevB.79.125443} {\bibfield
  {journal} {\bibinfo  {journal} {Phys. Rev. B}\ }\textbf {\bibinfo {volume}
  {79}},\ \bibinfo {pages} {125443} (\bibinfo {year} {2009})}\BibitemShut
  {NoStop}%
\bibitem [{\citenamefont {Wu}(2011)}]{wutrilayer}%
  \BibitemOpen
  \bibfield  {author} {\bibinfo {author} {\bibfnamefont {B.-R.}\ \bibnamefont
  {Wu}},\ }\href {\doibase 10.1063/1.3604019} {\bibfield  {journal} {\bibinfo
  {journal} {Appl. Phys. Lett.}\ }\textbf {\bibinfo {volume} {98}},\ \bibinfo
  {pages} {263107} (\bibinfo {year} {2011})}\BibitemShut {NoStop}%
\bibitem [{\citenamefont {Zhang}\ \emph {et~al.}(2010)\citenamefont {Zhang},
  \citenamefont {Sahu}, \citenamefont {Min},\ and\ \citenamefont
  {MacDonald}}]{zhangfantrilayer}%
  \BibitemOpen
  \bibfield  {author} {\bibinfo {author} {\bibfnamefont {F.}~\bibnamefont
  {Zhang}}, \bibinfo {author} {\bibfnamefont {B.}~\bibnamefont {Sahu}},
  \bibinfo {author} {\bibfnamefont {H.}~\bibnamefont {Min}}, \ and\ \bibinfo
  {author} {\bibfnamefont {A.~H.}\ \bibnamefont {MacDonald}},\ }\href {\doibase
  10.1103/PhysRevB.82.035409} {\bibfield  {journal} {\bibinfo  {journal} {Phys.
  Rev. B}\ }\textbf {\bibinfo {volume} {82}},\ \bibinfo {pages} {035409}
  (\bibinfo {year} {2010})}\BibitemShut {NoStop}%
\bibitem [{\citenamefont {Koshino}(2010)}]{koshino2010}%
  \BibitemOpen
  \bibfield  {author} {\bibinfo {author} {\bibfnamefont {M.}~\bibnamefont
  {Koshino}},\ }\href {\doibase 10.1103/PhysRevB.81.125304} {\bibfield
  {journal} {\bibinfo  {journal} {Phys. Rev. B}\ }\textbf {\bibinfo {volume}
  {81}},\ \bibinfo {pages} {125304} (\bibinfo {year} {2010})}\BibitemShut
  {NoStop}%
\bibitem [{\citenamefont {Tang}\ \emph {et~al.}(2011)\citenamefont {Tang},
  \citenamefont {Qin}, \citenamefont {Zhou}, \citenamefont {Qu}, \citenamefont
  {Zheng}, \citenamefont {Fei}, \citenamefont {Li}, \citenamefont {Zheng},
  \citenamefont {Gao},\ and\ \citenamefont {Lu}}]{lujingtrilayer}%
  \BibitemOpen
  \bibfield  {author} {\bibinfo {author} {\bibfnamefont {K.}~\bibnamefont
  {Tang}}, \bibinfo {author} {\bibfnamefont {R.}~\bibnamefont {Qin}}, \bibinfo
  {author} {\bibfnamefont {J.}~\bibnamefont {Zhou}}, \bibinfo {author}
  {\bibfnamefont {H.}~\bibnamefont {Qu}}, \bibinfo {author} {\bibfnamefont
  {J.}~\bibnamefont {Zheng}}, \bibinfo {author} {\bibfnamefont
  {R.}~\bibnamefont {Fei}}, \bibinfo {author} {\bibfnamefont {H.}~\bibnamefont
  {Li}}, \bibinfo {author} {\bibfnamefont {Q.}~\bibnamefont {Zheng}}, \bibinfo
  {author} {\bibfnamefont {Z.}~\bibnamefont {Gao}}, \ and\ \bibinfo {author}
  {\bibfnamefont {J.}~\bibnamefont {Lu}},\ }\href {\doibase 10.1021/jp201761p}
  {\bibfield  {journal} {\bibinfo  {journal} {J. Phys. Chem. C}\ }\textbf
  {\bibinfo {volume} {115}},\ \bibinfo {pages} {9458} (\bibinfo {year}
  {2011})}\BibitemShut {NoStop}%
\bibitem [{\citenamefont {Lui}\ \emph {et~al.}(2011)\citenamefont {Lui},
  \citenamefont {Li}, \citenamefont {Mak}, \citenamefont {Cappelluti},\ and\
  \citenamefont {Heinz}}]{heinz2011}%
  \BibitemOpen
  \bibfield  {author} {\bibinfo {author} {\bibfnamefont {C.~H.}\ \bibnamefont
  {Lui}}, \bibinfo {author} {\bibfnamefont {Z.}~\bibnamefont {Li}}, \bibinfo
  {author} {\bibfnamefont {K.~F.}\ \bibnamefont {Mak}}, \bibinfo {author}
  {\bibfnamefont {E.}~\bibnamefont {Cappelluti}}, \ and\ \bibinfo {author}
  {\bibfnamefont {T.~F.}\ \bibnamefont {Heinz}},\ }\href {\doibase
  10.1038/nphys2102} {\bibfield  {journal} {\bibinfo  {journal} {Nat. Phys.}\
  }\textbf {\bibinfo {volume} {7}},\ \bibinfo {pages} {944} (\bibinfo {year}
  {2011})}\BibitemShut {NoStop}%
\bibitem [{\citenamefont {Yankowitz}\ \emph {et~al.}(2013)\citenamefont
  {Yankowitz}, \citenamefont {Wang}, \citenamefont {Lau},\ and\ \citenamefont
  {LeRoy}}]{stmtrilayergap}%
  \BibitemOpen
  \bibfield  {author} {\bibinfo {author} {\bibfnamefont {M.}~\bibnamefont
  {Yankowitz}}, \bibinfo {author} {\bibfnamefont {F.}~\bibnamefont {Wang}},
  \bibinfo {author} {\bibfnamefont {C.~N.}\ \bibnamefont {Lau}}, \ and\
  \bibinfo {author} {\bibfnamefont {B.~J.}\ \bibnamefont {LeRoy}},\ }\href
  {\doibase 10.1103/PhysRevB.87.165102} {\bibfield  {journal} {\bibinfo
  {journal} {Phys. Rev. B}\ }\textbf {\bibinfo {volume} {87}},\ \bibinfo
  {pages} {165102} (\bibinfo {year} {2013})}\BibitemShut {NoStop}%
\bibitem [{\citenamefont {Khodkov}\ \emph {et~al.}(2015)\citenamefont
  {Khodkov}, \citenamefont {Khrapach}, \citenamefont {Craciun},\ and\
  \citenamefont {Russo}}]{abctransport}%
  \BibitemOpen
  \bibfield  {author} {\bibinfo {author} {\bibfnamefont {T.}~\bibnamefont
  {Khodkov}}, \bibinfo {author} {\bibfnamefont {I.}~\bibnamefont {Khrapach}},
  \bibinfo {author} {\bibfnamefont {M.~F.}\ \bibnamefont {Craciun}}, \ and\
  \bibinfo {author} {\bibfnamefont {S.}~\bibnamefont {Russo}},\ }\href
  {\doibase 10.1021/acs.nanolett.5b00772} {\bibfield  {journal} {\bibinfo
  {journal} {Nano Lett.}\ }\textbf {\bibinfo {volume} {15}},\ \bibinfo {pages}
  {4429} (\bibinfo {year} {2015})}\BibitemShut {NoStop}%
\bibitem [{\citenamefont {Xu}\ \emph {et~al.}(2015)\citenamefont {Xu},
  \citenamefont {Yin}, \citenamefont {Qiao}, \citenamefont {Bai}, \citenamefont
  {Nie},\ and\ \citenamefont {He}}]{helin2015}%
  \BibitemOpen
  \bibfield  {author} {\bibinfo {author} {\bibfnamefont {R.}~\bibnamefont
  {Xu}}, \bibinfo {author} {\bibfnamefont {L.-J.}\ \bibnamefont {Yin}},
  \bibinfo {author} {\bibfnamefont {J.-B.}\ \bibnamefont {Qiao}}, \bibinfo
  {author} {\bibfnamefont {K.-K.}\ \bibnamefont {Bai}}, \bibinfo {author}
  {\bibfnamefont {J.-C.}\ \bibnamefont {Nie}}, \ and\ \bibinfo {author}
  {\bibfnamefont {L.}~\bibnamefont {He}},\ }\href {\doibase
  10.1103/PhysRevB.91.035410} {\bibfield  {journal} {\bibinfo  {journal} {Phys.
  Rev. B}\ }\textbf {\bibinfo {volume} {91}},\ \bibinfo {pages} {035410}
  (\bibinfo {year} {2015})}\BibitemShut {NoStop}%
\bibitem [{\citenamefont {Ding}\ \emph {et~al.}(2009)\citenamefont {Ding},
  \citenamefont {Zhu},\ and\ \citenamefont {Berakdar}}]{ding2009}%
  \BibitemOpen
  \bibfield  {author} {\bibinfo {author} {\bibfnamefont {K.-H.}\ \bibnamefont
  {Ding}}, \bibinfo {author} {\bibfnamefont {Z.-G.}\ \bibnamefont {Zhu}}, \
  and\ \bibinfo {author} {\bibfnamefont {J.}~\bibnamefont {Berakdar}},\ }\href
  {http://stacks.iop.org/0953-8984/21/i=18/a=182002} {\bibfield  {journal}
  {\bibinfo  {journal} {J. Phys.: Condens. Matter}\ }\textbf {\bibinfo {volume}
  {21}},\ \bibinfo {pages} {182002} (\bibinfo {year} {2009})}\BibitemShut
  {NoStop}%
\bibitem [{\citenamefont {Zhang}\ \emph {et~al.}(2009)\citenamefont {Zhang},
  \citenamefont {Tang}, \citenamefont {Girit}, \citenamefont {Hao},
  \citenamefont {Martin}, \citenamefont {Zettl}, \citenamefont {Crommie},
  \citenamefont {Shen},\ and\ \citenamefont {Wang}}]{zhangyuanbo2009}%
  \BibitemOpen
  \bibfield  {author} {\bibinfo {author} {\bibfnamefont {Y.}~\bibnamefont
  {Zhang}}, \bibinfo {author} {\bibfnamefont {T.-T.}\ \bibnamefont {Tang}},
  \bibinfo {author} {\bibfnamefont {C.}~\bibnamefont {Girit}}, \bibinfo
  {author} {\bibfnamefont {Z.}~\bibnamefont {Hao}}, \bibinfo {author}
  {\bibfnamefont {M.~C.}\ \bibnamefont {Martin}}, \bibinfo {author}
  {\bibfnamefont {A.}~\bibnamefont {Zettl}}, \bibinfo {author} {\bibfnamefont
  {M.~F.}\ \bibnamefont {Crommie}}, \bibinfo {author} {\bibfnamefont {Y.~R.}\
  \bibnamefont {Shen}}, \ and\ \bibinfo {author} {\bibfnamefont
  {F.}~\bibnamefont {Wang}},\ }\href {\doibase 10.1038/nature08105} {\bibfield
  {journal} {\bibinfo  {journal} {Nature}\ }\textbf {\bibinfo {volume} {459}},\
  \bibinfo {pages} {820} (\bibinfo {year} {2009})}\BibitemShut {NoStop}%
\bibitem [{\citenamefont {Lee}\ \emph {et~al.}(2013)\citenamefont {Lee},
  \citenamefont {Velasco}, \citenamefont {Tran}, \citenamefont {Zhang},
  \citenamefont {Bao}, \citenamefont {Jing}, \citenamefont {Myhro},
  \citenamefont {Smirnov},\ and\ \citenamefont {Lau}}]{lauaba}%
  \BibitemOpen
  \bibfield  {author} {\bibinfo {author} {\bibfnamefont {Y.}~\bibnamefont
  {Lee}}, \bibinfo {author} {\bibfnamefont {J.}~\bibnamefont {Velasco}},
  \bibinfo {author} {\bibfnamefont {D.}~\bibnamefont {Tran}}, \bibinfo {author}
  {\bibfnamefont {F.}~\bibnamefont {Zhang}}, \bibinfo {author} {\bibfnamefont
  {W.}~\bibnamefont {Bao}}, \bibinfo {author} {\bibfnamefont {L.}~\bibnamefont
  {Jing}}, \bibinfo {author} {\bibfnamefont {K.}~\bibnamefont {Myhro}},
  \bibinfo {author} {\bibfnamefont {D.}~\bibnamefont {Smirnov}}, \ and\
  \bibinfo {author} {\bibfnamefont {C.~N.}\ \bibnamefont {Lau}},\ }\href
  {\doibase 10.1021/nl4000757} {\bibfield  {journal} {\bibinfo  {journal} {Nano
  Lett.}\ }\textbf {\bibinfo {volume} {13}},\ \bibinfo {pages} {1627} (\bibinfo
  {year} {2013})}\BibitemShut {NoStop}%
\bibitem [{\citenamefont {Lee}\ \emph {et~al.}(2014)\citenamefont {Lee},
  \citenamefont {Tran}, \citenamefont {Myhro}, \citenamefont {Velasco},
  \citenamefont {Gillgren}, \citenamefont {Lau}, \citenamefont {Barlas},
  \citenamefont {Poumirol}, \citenamefont {Smirnov},\ and\ \citenamefont
  {Guinea}}]{lau2014}%
  \BibitemOpen
  \bibfield  {author} {\bibinfo {author} {\bibfnamefont {Y.}~\bibnamefont
  {Lee}}, \bibinfo {author} {\bibfnamefont {D.}~\bibnamefont {Tran}}, \bibinfo
  {author} {\bibfnamefont {K.}~\bibnamefont {Myhro}}, \bibinfo {author}
  {\bibfnamefont {J.}~\bibnamefont {Velasco}}, \bibinfo {author} {\bibfnamefont
  {N.}~\bibnamefont {Gillgren}}, \bibinfo {author} {\bibfnamefont {C.~N.}\
  \bibnamefont {Lau}}, \bibinfo {author} {\bibfnamefont {Y.}~\bibnamefont
  {Barlas}}, \bibinfo {author} {\bibfnamefont {J.~M.}\ \bibnamefont
  {Poumirol}}, \bibinfo {author} {\bibfnamefont {D.}~\bibnamefont {Smirnov}}, \
  and\ \bibinfo {author} {\bibfnamefont {F.}~\bibnamefont {Guinea}},\ }\href
  {\doibase 10.1038/ncomms6656} {\bibfield  {journal} {\bibinfo  {journal}
  {Nat. Commun.}\ }\textbf {\bibinfo {volume} {5}},\ \bibinfo {pages} {5656}
  (\bibinfo {year} {2014})}\BibitemShut {NoStop}%
\bibitem [{\citenamefont {Zhang}\ \emph {et~al.}(2011)\citenamefont {Zhang},
  \citenamefont {Lin}, \citenamefont {Liu}, \citenamefont {Tite}, \citenamefont
  {Su}, \citenamefont {Chang}, \citenamefont {Lee}, \citenamefont {Chu},
  \citenamefont {Wei}, \citenamefont {Kuo},\ and\ \citenamefont
  {Li}}]{dopinggap}%
  \BibitemOpen
  \bibfield  {author} {\bibinfo {author} {\bibfnamefont {W.}~\bibnamefont
  {Zhang}}, \bibinfo {author} {\bibfnamefont {C.-T.}\ \bibnamefont {Lin}},
  \bibinfo {author} {\bibfnamefont {K.-K.}\ \bibnamefont {Liu}}, \bibinfo
  {author} {\bibfnamefont {T.}~\bibnamefont {Tite}}, \bibinfo {author}
  {\bibfnamefont {C.-Y.}\ \bibnamefont {Su}}, \bibinfo {author} {\bibfnamefont
  {C.-H.}\ \bibnamefont {Chang}}, \bibinfo {author} {\bibfnamefont {Y.-H.}\
  \bibnamefont {Lee}}, \bibinfo {author} {\bibfnamefont {C.-W.}\ \bibnamefont
  {Chu}}, \bibinfo {author} {\bibfnamefont {K.-H.}\ \bibnamefont {Wei}},
  \bibinfo {author} {\bibfnamefont {J.-L.}\ \bibnamefont {Kuo}}, \ and\
  \bibinfo {author} {\bibfnamefont {L.-J.}\ \bibnamefont {Li}},\ }\href
  {\doibase 10.1021/nn202463g} {\bibfield  {journal} {\bibinfo  {journal} {ACS
  Nano}\ }\textbf {\bibinfo {volume} {5}},\ \bibinfo {pages} {7517} (\bibinfo
  {year} {2011})}\BibitemShut {NoStop}%
\bibitem [{\citenamefont {Yu}\ \emph {et~al.}(2011)\citenamefont {Yu},
  \citenamefont {Liao}, \citenamefont {Chae}, \citenamefont {Lee},\ and\
  \citenamefont {Duan}}]{duan2011}%
  \BibitemOpen
  \bibfield  {author} {\bibinfo {author} {\bibfnamefont {W.~J.}\ \bibnamefont
  {Yu}}, \bibinfo {author} {\bibfnamefont {L.}~\bibnamefont {Liao}}, \bibinfo
  {author} {\bibfnamefont {S.~H.}\ \bibnamefont {Chae}}, \bibinfo {author}
  {\bibfnamefont {Y.~H.}\ \bibnamefont {Lee}}, \ and\ \bibinfo {author}
  {\bibfnamefont {X.}~\bibnamefont {Duan}},\ }\href {\doibase
  10.1021/nl2025739} {\bibfield  {journal} {\bibinfo  {journal} {Nano Lett.}\
  }\textbf {\bibinfo {volume} {11}},\ \bibinfo {pages} {4759} (\bibinfo {year}
  {2011})}\BibitemShut {NoStop}%
\bibitem [{\citenamefont {Han}\ \emph {et~al.}(2015)\citenamefont {Han},
  \citenamefont {Yan}, \citenamefont {Jia}, \citenamefont {Niu}, \citenamefont
  {Yu},\ and\ \citenamefont {Wu}}]{wuxiaosong}%
  \BibitemOpen
  \bibfield  {author} {\bibinfo {author} {\bibfnamefont {Q.}~\bibnamefont
  {Han}}, \bibinfo {author} {\bibfnamefont {B.}~\bibnamefont {Yan}}, \bibinfo
  {author} {\bibfnamefont {Z.}~\bibnamefont {Jia}}, \bibinfo {author}
  {\bibfnamefont {J.}~\bibnamefont {Niu}}, \bibinfo {author} {\bibfnamefont
  {D.}~\bibnamefont {Yu}}, \ and\ \bibinfo {author} {\bibfnamefont
  {X.}~\bibnamefont {Wu}},\ }\href {\doibase 10.1063/1.4934489} {\bibfield
  {journal} {\bibinfo  {journal} {Appl. Phys. Lett.}\ }\textbf {\bibinfo
  {volume} {107}},\ \bibinfo {pages} {163104} (\bibinfo {year}
  {2015})}\BibitemShut {NoStop}%
\bibitem [{\citenamefont {Killi}\ \emph
  {et~al.}(2011{\natexlab{a}})\citenamefont {Killi}, \citenamefont
  {Heidarian},\ and\ \citenamefont {Paramekanti}}]{levelshift}%
  \BibitemOpen
  \bibfield  {author} {\bibinfo {author} {\bibfnamefont {M.}~\bibnamefont
  {Killi}}, \bibinfo {author} {\bibfnamefont {D.}~\bibnamefont {Heidarian}}, \
  and\ \bibinfo {author} {\bibfnamefont {A.}~\bibnamefont {Paramekanti}},\
  }\href {\doibase 10.1088/1367-2630/13/5/053043} {\bibfield  {journal}
  {\bibinfo  {journal} {New J. Phys.}\ }\textbf {\bibinfo {volume} {13}},\
  \bibinfo {pages} {053043} (\bibinfo {year} {2011}{\natexlab{a}})}\BibitemShut
  {NoStop}%
\bibitem [{\citenamefont {Que}\ \emph {et~al.}(2015)\citenamefont {Que},
  \citenamefont {Xiao}, \citenamefont {Chen}, \citenamefont {Wang},
  \citenamefont {Du},\ and\ \citenamefont {Gao}}]{xiao2015}%
  \BibitemOpen
  \bibfield  {author} {\bibinfo {author} {\bibfnamefont {Y.}~\bibnamefont
  {Que}}, \bibinfo {author} {\bibfnamefont {W.}~\bibnamefont {Xiao}}, \bibinfo
  {author} {\bibfnamefont {H.}~\bibnamefont {Chen}}, \bibinfo {author}
  {\bibfnamefont {D.}~\bibnamefont {Wang}}, \bibinfo {author} {\bibfnamefont
  {S.}~\bibnamefont {Du}}, \ and\ \bibinfo {author} {\bibfnamefont {H.-J.}\
  \bibnamefont {Gao}},\ }\href {\doibase 10.1063/1.4938466} {\bibfield
  {journal} {\bibinfo  {journal} {Appl. Phys. Lett.}\ }\textbf {\bibinfo
  {volume} {107}},\ \bibinfo {pages} {263101} (\bibinfo {year}
  {2015})}\BibitemShut {NoStop}%
\bibitem [{\citenamefont {Luo}\ \emph {et~al.}(2004)\citenamefont {Luo},
  \citenamefont {Xiang}, \citenamefont {Wang}, \citenamefont {Su},\ and\
  \citenamefont {Yu}}]{hgluo2004}%
  \BibitemOpen
  \bibfield  {author} {\bibinfo {author} {\bibfnamefont {H.~G.}\ \bibnamefont
  {Luo}}, \bibinfo {author} {\bibfnamefont {T.}~\bibnamefont {Xiang}}, \bibinfo
  {author} {\bibfnamefont {X.~Q.}\ \bibnamefont {Wang}}, \bibinfo {author}
  {\bibfnamefont {Z.~B.}\ \bibnamefont {Su}}, \ and\ \bibinfo {author}
  {\bibfnamefont {L.}~\bibnamefont {Yu}},\ }\href {\doibase
  10.1103/PhysRevLett.92.256602} {\bibfield  {journal} {\bibinfo  {journal}
  {Phys. Rev. Lett.}\ }\textbf {\bibinfo {volume} {92}},\ \bibinfo {pages}
  {256602} (\bibinfo {year} {2004})}\BibitemShut {NoStop}%
\bibitem [{\citenamefont {Nilsson}\ and\ \citenamefont
  {Castro~Neto}(2007)}]{ingap2007}%
  \BibitemOpen
  \bibfield  {author} {\bibinfo {author} {\bibfnamefont {J.}~\bibnamefont
  {Nilsson}}\ and\ \bibinfo {author} {\bibfnamefont {A.~H.}\ \bibnamefont
  {Castro~Neto}},\ }\href {\doibase 10.1103/PhysRevLett.98.126801} {\bibfield
  {journal} {\bibinfo  {journal} {Phys. Rev. Lett.}\ }\textbf {\bibinfo
  {volume} {98}},\ \bibinfo {pages} {126801} (\bibinfo {year}
  {2007})}\BibitemShut {NoStop}%
\bibitem [{\citenamefont {Castro}\ \emph {et~al.}(2010)\citenamefont {Castro},
  \citenamefont {L\'opez-Sancho},\ and\ \citenamefont
  {Vozmediano}}]{vacancy2010}%
  \BibitemOpen
  \bibfield  {author} {\bibinfo {author} {\bibfnamefont {E.~V.}\ \bibnamefont
  {Castro}}, \bibinfo {author} {\bibfnamefont {M.~P.}\ \bibnamefont
  {L\'opez-Sancho}}, \ and\ \bibinfo {author} {\bibfnamefont {M.~A.~H.}\
  \bibnamefont {Vozmediano}},\ }\href {\doibase 10.1103/PhysRevLett.104.036802}
  {\bibfield  {journal} {\bibinfo  {journal} {Phys. Rev. Lett.}\ }\textbf
  {\bibinfo {volume} {104}},\ \bibinfo {pages} {036802} (\bibinfo {year}
  {2010})}\BibitemShut {NoStop}%
\bibitem [{\citenamefont {Mkhitaryan}\ and\ \citenamefont
  {Mishchenko}(2013)}]{ingap2013}%
  \BibitemOpen
  \bibfield  {author} {\bibinfo {author} {\bibfnamefont {V.~V.}\ \bibnamefont
  {Mkhitaryan}}\ and\ \bibinfo {author} {\bibfnamefont {E.~G.}\ \bibnamefont
  {Mishchenko}},\ }\href {\doibase 10.1103/PhysRevLett.110.086805} {\bibfield
  {journal} {\bibinfo  {journal} {Phys. Rev. Lett.}\ }\textbf {\bibinfo
  {volume} {110}},\ \bibinfo {pages} {086805} (\bibinfo {year}
  {2013})}\BibitemShut {NoStop}%
\bibitem [{\citenamefont {Ojeda~Collado}\ \emph {et~al.}(2015)\citenamefont
  {Ojeda~Collado}, \citenamefont {Usaj},\ and\ \citenamefont
  {Balseiro}}]{ingap2015}%
  \BibitemOpen
  \bibfield  {author} {\bibinfo {author} {\bibfnamefont {H.~P.}\ \bibnamefont
  {Ojeda~Collado}}, \bibinfo {author} {\bibfnamefont {G.}~\bibnamefont {Usaj}},
  \ and\ \bibinfo {author} {\bibfnamefont {C.~A.}\ \bibnamefont {Balseiro}},\
  }\href {\doibase 10.1103/PhysRevB.91.045435} {\bibfield  {journal} {\bibinfo
  {journal} {Phys. Rev. B}\ }\textbf {\bibinfo {volume} {91}},\ \bibinfo
  {pages} {045435} (\bibinfo {year} {2015})}\BibitemShut {NoStop}%
\bibitem [{\citenamefont {Dahal}\ \emph {et~al.}(2008)\citenamefont {Dahal},
  \citenamefont {Balatsky},\ and\ \citenamefont {Zhu}}]{zhujianxin}%
  \BibitemOpen
  \bibfield  {author} {\bibinfo {author} {\bibfnamefont {H.~P.}\ \bibnamefont
  {Dahal}}, \bibinfo {author} {\bibfnamefont {A.~V.}\ \bibnamefont {Balatsky}},
  \ and\ \bibinfo {author} {\bibfnamefont {J.-X.}\ \bibnamefont {Zhu}},\ }\href
  {\doibase 10.1103/PhysRevB.77.115114} {\bibfield  {journal} {\bibinfo
  {journal} {Phys. Rev. B}\ }\textbf {\bibinfo {volume} {77}},\ \bibinfo
  {pages} {115114} (\bibinfo {year} {2008})}\BibitemShut {NoStop}%
\bibitem [{\citenamefont {Killi}\ \emph
  {et~al.}(2011{\natexlab{b}})\citenamefont {Killi}, \citenamefont
  {Heidarian},\ and\ \citenamefont {Paramekanti}}]{bilayer2011}%
  \BibitemOpen
  \bibfield  {author} {\bibinfo {author} {\bibfnamefont {M.}~\bibnamefont
  {Killi}}, \bibinfo {author} {\bibfnamefont {D.}~\bibnamefont {Heidarian}}, \
  and\ \bibinfo {author} {\bibfnamefont {A.}~\bibnamefont {Paramekanti}},\
  }\href {\doibase 10.1088/1367-2630/13/5/053043} {\bibfield  {journal}
  {\bibinfo  {journal} {New J. Phys.}\ }\textbf {\bibinfo {volume} {13}},\
  \bibinfo {pages} {053043} (\bibinfo {year} {2011}{\natexlab{b}})}\BibitemShut
  {NoStop}%
\bibitem [{\citenamefont {Mohammadi}\ and\ \citenamefont
  {Moradian}(2014{\natexlab{a}})}]{bilayer2014}%
  \BibitemOpen
  \bibfield  {author} {\bibinfo {author} {\bibfnamefont {Y.}~\bibnamefont
  {Mohammadi}}\ and\ \bibinfo {author} {\bibfnamefont {R.}~\bibnamefont
  {Moradian}},\ }\href {\doibase http://dx.doi.org/10.1016/j.ssc.2013.10.021}
  {\bibfield  {journal} {\bibinfo  {journal} {Solid State Commun.}\ }\textbf
  {\bibinfo {volume} {178}},\ \bibinfo {pages} {37 } (\bibinfo {year}
  {2014}{\natexlab{a}})}\BibitemShut {NoStop}%
\bibitem [{\citenamefont {Mohammadi}\ and\ \citenamefont
  {Moradian}(2014{\natexlab{b}})}]{aabilayer}%
  \BibitemOpen
  \bibfield  {author} {\bibinfo {author} {\bibfnamefont {Y.}~\bibnamefont
  {Mohammadi}}\ and\ \bibinfo {author} {\bibfnamefont {R.}~\bibnamefont
  {Moradian}},\ }\href {\doibase http://dx.doi.org/10.1016/j.physb.2014.02.010}
  {\bibfield  {journal} {\bibinfo  {journal} {Physica B}\ }\textbf {\bibinfo
  {volume} {442}},\ \bibinfo {pages} {66 } (\bibinfo {year}
  {2014}{\natexlab{b}})}\BibitemShut {NoStop}%
\end{thebibliography}%

\end{document}